\documentclass[a4paper,11pt]{article}
\pdfoutput=1 

\usepackage{jheppub} 
\usepackage{amsmath,amssymb,amsthm,amscd,graphicx}
\usepackage{slashed}
\usepackage[notref,notcite]{}
\input epsf.sty

\theoremstyle{definition}

\newcommand{\CB}{{\cal B}}
\newcommand{\CC}{{\cal C}}

\newcommand{\CF}{{\cal F}}

\newcommand{\CI}{{\cal I}}
\newcommand{\CJ}{{\cal J}}

\newcommand{\CL}{{\cal L}}

\newcommand{\CN}{{\cal N}}
\newcommand{\CO}{{\cal O}}

\newcommand{\CS}{{\cal S}}

\def\IR{{\mathbb R}}

\def\bS{\boldsymbol{S}}

\newcommand{\mO}{\mathsf{O}}

\newcommand{\mQ}{\mathsf{Q}}

\newcommand{\mH}{\mathsf{H}}

\newcommand{\e}{{\rm e}}
\def\gbar{{\overline{g}}}
\newcommand{\tr}{{\rm Tr}}
\newcommand{\re}{{\rm e}}
\newcommand{\ri}{{\rm i}}
\newcommand{\rd}{{\rm d}}

\newcommand{\be}{\begin{equation}}
\newcommand{\ee}{\end{equation}}
\newcommand{\ba}{\begin{aligned}}
\newcommand{\ea}{\end{aligned}}
\newcommand{\ben}{\begin{eqnarray}\displaystyle}
\newcommand{\een}{\end{eqnarray}}

\newcommand{\sectiono}[1]{\section{#1}\setcounter{equation}{0}}

\newdimen\tableauside\tableauside=1.0ex
\newdimen\tableaurule\tableaurule=0.4pt
\newdimen\tableaustep
\def\phantomhrule#1{\hbox{\vbox to0pt{\hrule height\tableaurule width#1\vss}}}
\def\phantomvrule#1{\vbox{\hbox to0pt{\vrule width\tableaurule height#1\hss}}}
\def\sqr{\vbox{%
  \phantomhrule\tableaustep
  \hbox{\phantomvrule\tableaustep\kern\tableaustep\phantomvrule\tableaustep}%
  \hbox{\vbox{\phantomhrule\tableauside}\kern-\tableaurule}}}
\def\squares#1{\hbox{\count0=#1\noindent\loop\sqr
  \advance\count0 by-1 \ifnum\count0>0\repeat}}
\def\tableau#1{\vcenter{\offinterlineskip
  \tableaustep=\tableauside\advance\tableaustep by-\tableaurule
  \kern\normallineskip\hbox
    {\kern\normallineskip\vbox
      {\gettableau#1 0 }%
     \kern\normallineskip\kern\tableaurule}%
  \kern\normallineskip\kern\tableaurule}}
\def\gettableau#1{\ifnum#1=0\let\next=\null\else
\squares{#1}\let\next=\gettableau\fi\next}

\newcommand{\figref}[1]{Fig.~\protect\ref{#1}}

\title{\Huge{\boldmath Renormalons in integrable field theories}}

\author{Marcos Mari\~no and Tom\'as Reis}

\affiliation{D\'epartement de Physique Th\'eorique et Section de Math\'ematiques\\
Universit\'e de Gen\`eve, Gen\`eve, CH-1211 Switzerland}

\abstract{In integrable field theories in two dimensions, the Bethe ansatz can be used to compute exactly the ground state energy in the presence of 
an external field coupled to a conserved charge. We generalize previous results by Volin and we extract analytic results for the 
perturbative expansion of this observable, up to very high order, in various asymptotically free theories: the non-linear sigma model and 
its supersymmetric extension, the Gross--Neveu model, and the principal chiral field. 
We study the large order behavior of these perturbative series and we give strong evidence that, as expected, 
it is controlled by renormalons. Our analysis is sensitive to the next-to-leading correction to the asymptotics, which involves the first two coefficients 
of the beta function. }

\begin{document}
\maketitle
\flushbottom

\sectiono{Introduction}

Understanding the large order behavior of perturbative series in quantum theory is a possible route to 
unveiling non-perturbative effects. In quantum mechanics and in many 
super-renormalizable field theories, the coefficients of the perturbative series grow factorially, due to the growth of the total number of 
diagrams with the number of loops \cite{bw-stat}. This behavior is controlled by instantons \cite{lam,bw2}, and therefore it has a semiclassical 
description. The relation 
between instantons and large-order behavior has led to many beautiful results and it has evolved into the theory of resurgence, which provides a 
universal structure linking perturbative and non-perturbative sectors in quantum theories (see e.g. \cite{zjbook, mmbook} for a 
presentation of instanton-induced large order behavior, and \cite{mmlargen, abs} for reviews of the theory of resurgence). 

However, in the 1970s it was found that, in renormalizable field theories, the large order 
behavior of the perturbative series involves a different type of phenomenon 
\cite{gross-neveu,lautrup,parisi1,parisi2, thooft}: one can find specific diagrams which grow factorially 
with the loop order after integration over the momenta\footnote{According to standard lore, 
renormalons, as their name indicate, appear only in renormalizable field theories. However, 
renormalon behavior has been recently found in condensed matter systems \cite{mr-long, mr-ll}, 
in quantum mechanics \cite{rqm}, and somewhat surprisingly, in some 
super-renormalizable theories \cite{mr-rtd}.}. These diagrams are 
usually called {\it renormalon diagrams} (see \cite{beneke} for an extensive review and \cite{shifman} for 
an invitation to the subject). They lead to singularities in the Borel plane of the coupling constant which, following \cite{beneke}, we will call 
renormalon singularities, or renormalons for short. Depending on the region 
in momenta which leads to the factorial growth, one has UV or IR renormalons. The analysis of renormalon diagrams is 
mostly based on heuristic and plausibility arguments. It is relatively easy to find sequences of diagrams with the appropriate factorial growth, like 
the famous bubble chain diagrams in QED and QCD. However, the resulting behavior could be in principle corrected or even 
cancelled by some other set of diagrams. Usually, the 
identification of renormalon diagrams is combined with some type of large $N$ limit (here $N$ can be the number of components of a 
field in the theory, or the number of fermions, or any other convenient counting parameter), so that one can at least argue that renormalon effects 
appear unequivocally at large $N$. In cases where the operator product expansion (OPE) is available, IR renormalon effects 
can be shown to correspond to non-perturbative condensates in the OPE \cite{parisi2}, and this is usually 
regarded as evidence for both IR renormalon physics and the validity of the OPE. 

Renormalon effects are believed to control the large order behavior of 
perturbative series in many renormalizable theories, as they are typically more important than instanton effects. Therefore, the cleanest way of establishing the 
presence of renormalon effects is to show explicitly that the perturbative 
series has the asymptotic behavior dictated by them. However, it is 
in general difficult to produce explicit values for a large number of 
coefficients in the series, so this type of tests are difficult to make. A notable exception 
is the {\it tour de force} numerical computation in 
\cite{pineda, pineda2}, which gives 
a beautiful and precise test of renormalon predictions in Yang--Mills theory. 

Given the subtleties of renormalon physics, it is useful to look at simple field theories where one has more 
analytic control. For example, renormalons in the 
non-linear sigma model at large $N$ were analyzed in some detail in 
\cite{david1,david2, david3,itep,beneke-braun}, and very recently evidence for the dominance of the leading IR 
renormalon was obtained numerically in \cite{pcf-lattice}, in the 
case of the principal chiral field (PCF) \cite{polyakov}. 
Many asymptotically free 
theories in two dimensions turn out to be integrable, i.e. the $S$-matrix is known exactly. The 
Bethe ansatz can be then used to compute the free energy of these theories in the presence of an external field coupled to 
a conserved current \cite{pw}. This has made it possible to obtain the exact mass gap of these theories in various cases 
\cite{hmn,hn,pcf,fnw1,fnw2,h-pcf,eh-ssm,eh-scpn}  (see \cite{eh-review} for a review). It was shown by Volin in \cite{volin, volin-thesis} that one 
can extend the mass gap calculation and extract from the Bethe ansatz 
the full perturbative series for the vacuum energy, as a function of the running coupling constant. Volin worked out the example of the 
non-linear sigma model, where he addressed some aspects of the large order behavior of the perturbative series. Related work appeared before in 
\cite{fkw1,fkw2}, where the PCF was analyzed, but with a different choice of conserved charge than what is made in \cite{pcf}. The leading large $N$ contribution 
to the ground state energy was obtained at all orders in the coupling constant, and it was noted that the resulting factorial divergence is due to renormalon effects. 

In this paper we generalize the results of \cite{volin, volin-thesis} in two directions. First of all, we streamline the method of resolution of the integral equation 
by combining it with the Wiener--Hopf method, as we already did in \cite{mr-long}. This makes it possible to 
obtain the perturbative series for the vacuum energy up 
to very large order in various integrable models, namely, the supersymmetric $O(N)$ non-linear sigma model \cite{witten}, the 
$SU(N)$ PCF \cite{polyakov}, and the $O(N)$ Gross--Neveu (GN) model \cite{gross-neveu}, in all cases for arbitrary finite $N$. 
Second, we do a precision analysis of the resulting perturbative series in 
order to test the predictions of renormalon physics. Care is needed since the large 
order behavior mixes IR and UV renormalons, and one needs to disentangle their contributions to the 
asymptotics. As a result, we are able to 
test the predictions of renormalon physics to next-to-leading order in the asymptotics. This subleading correction involves the two coefficients 
of the beta function of the theory. 

The organization of this paper is as follows. In section \ref{pseries-sec} we 
review the theories that we will analyze, the observables that we want to compute, 
and we explain how to 
extract perturbative series from the Bethe ansatz. In section \ref{ren-sec} we review the 
predictions or renormalon physics for the large order behaviour of these perturbative series. 
Then, we compare these expectations to our data. Finally, we conclude with some open 
problems raised by this investigation.

\sectiono{Perturbative series from integrability}

\label{pseries-sec}

\subsection{Integrable asymptotically free theories in two dimensions}

In this paper we will consider integrable quantum field theories in two dimensions which are also asymptotically free. Our convention for the beta function is 
\be
\label{betaf}
\beta(g)= \mu {\rd g \over \rd \mu} =-\beta_0 g^3 - \beta_1 g^5- \cdots, 
\ee
and we will denote 
\be
\label{xi-constant}
\xi = \frac{\beta_1}{2 \beta_0^2}.
\ee
With the convention above, asymptotically free theories have $\beta_0>0$. 

We will consider two types of theories: the ``bosonic" theories include the non-linear $O(N)$ sigma model, its $\CN=1$ supersymmetrix extension, and the $SU(N)$ PCF. The ``fermionic" theory will be the $O(N)$ Gross--Neveu model. Let $\mH$ the Hamiltonian of any of these theories, $\mQ$ the charge associated to a global conserved 
current, and $h$ an external field coupled to $\mQ$. The external field $h$ can be regarded as a chemical potential, and as usual in statistical mechanics we can consider the ensemble defined by the operator
\be
\label{HQ}
\mH- h \mQ.
\ee
The corresponding free energy per unit volume is then defined by 
\be
\label{free-en}
\CF(h) =-\lim_{V, \beta \rightarrow \infty} {1\over V\beta } \log \, \tr \, \re^{-\beta (\mH-h \mQ)}, 
\ee
where $V$ is the volume of space and $\beta$ is the total length of Euclidean time. 

When $h$ is large, and since the theories we are considering are asymptotically free, one can calculate $\CF(h)$ in perturbation theory. 
We can use the renormalization group (RG) to re-express the perturbative series in terms of the RG-invariant coupling $\overline g^2 (\mu/h, g)$, defined by 
\be
\label{RG-evol}
\log\left( {\mu \over h} \right)= -\int_{g}^{{\overline g}} {\rd x \over \beta(x)}. 
\ee
$\overline g$ can be expressed in terms of $\Lambda/h$, where $\Lambda$ is the dynamically generated scale, which we define as  
\be
\label{Lambda}
\Lambda=\mu \left( 2 \beta_0 g^2 \right)^{-\beta_1/(2 \beta_0^2)} \re^{-1/(2 \beta_0 g^2)} 
\exp\left( -\int_0^g \left\{ {1\over \beta(x)}+ {1\over \beta_0 x^3} -{\beta_1 \over \beta_0^2 x} \right\} \rd x  \right).
\ee
At leading order we have, 
\be
 {1\over {\overline g}^2}=2 \beta_0\left(\log\left(\frac{h}{\Lambda}\right)+\xi \log \log \left(\frac{h}{\Lambda}\right)\right)+ \cdots. 
 \ee

The theories we will consider are also integrable, and their $S$ matrix is known exactly. It was shown in \cite{pw} that this makes it possible to calculate the free energy 
(\ref{free-en}) by using the Bethe ansatz. After turning on the chemical potential $h$ beyond an appropriate 
threshold, there will be a density $\rho$ of particles charged under the conserved charge $\mQ$, with an energy per unit volume given by $e(\rho)$. These two quantities can be obtained from the density of Bethe roots $\chi(\theta)$. This density is supported on an interval $[-B,B]$ and satisfies the integral equation
\be
\label{chi-ie}
m \cosh \theta=\chi(\theta)-\int_{-B}^B  \rd \theta' \, K(\theta-\theta') \chi (\theta').
\ee
In this equation, $m$ is the mass of the charged particles, and with a clever choice of $\mQ$, it is directly related to the mass gap of the theory. 
The kernel of the integral equation is given by 
\be
K(\theta)={1\over 2 \pi \ri} {\rd \over \rd\theta} \log S(\theta),
\ee
where $S(\theta)$ is the $S$-matrix appropriate for the scattering of the charged particles. The energy per unit volume and the density are then given by
\be
\label{erho}
e={m \over 2 \pi} \int_{-B}^B \rd \theta \, \chi(\theta)  \cosh \theta, \qquad \rho={1\over 2 \pi} \int_{-B}^B \rd \theta\, \chi(\theta). 
\ee
Finally, the free energy $\CF(h)$ can be obtained as a Legendre transform of $e(\rho)$:
\be
\ba
\rho&=-\CF'(h), \\
\mathcal{F}(h)-\mathcal{F}(0)&=e(\rho)-\rho h.
\ea
\ee
Note that the first equation defines $\rho$ as a function of $h$. 

Integrable asymptotically free theories in two dimensions have been a useful laboratory to test general expectations from QFT. For example, 
in asymptotically free massless theories, the masses of the particles in the spectrum are expected to be proportional to the dynamically generated 
scale $\Lambda$, but the calculation of the proportionality constant is a difficult non-perturbative problem. It was noted in \cite{hmn,hn} that, in integrable models, this 
constant can be calculated exactly. The reason is as follows: in the calculation of $\CF(h)$ from the Bethe ansatz, the answer is naturally 
expressed in terms of $m/h$, where $m$ is the mass of the charged  particles under $\mQ$. 
On the other hand, the perturbative calculation gives the answer in terms of $\Lambda/h$. By matching these two expressions, one 
can find an exact expression for the physical mass as a function of $\Lambda$. This typically requires just a one-loop calculation in the ``bosonic" theories and 
a two-loop calculation in the GN model. The original calculation of \cite{hmn,hn} was done for the non-linear sigma model, but it was quickly 
generalized to the PCF \cite{pcf,h-pcf}, the Gross--Neveu model \cite{fnw1,fnw2}, and to supersymmetric models \cite{eh-ssm,eh-scpn} 
(see \cite{eh-review} for a review). In order to perform these computations, one has 
to solve the integral equation (\ref{chi-ie}) for large $h$, which corresponds to large $B$. This is technically 
challenging and it was done in \cite{hmn,hn,pcf,fnw1,fnw2,h-pcf,eh-ssm,eh-scpn} by using the Wiener--Hopf method. In the next section, following \cite{volin}, we will present 
a more powerful method to solve the integral equation, which can be used to generate the full perturbative series for $\CF(h)$, or, equivalently, for the ground state energy $e(\rho)$. 

Before doing this, let us list some of the basic ingredients in the four theories that we will study. 

\vskip .2cm

(i) {\it Non-linear $O(N)$ sigma model}. The basic field of the non-linear sigma model is a scalar 
field $\bS: \IR^2 \rightarrow \IR^N$ satisfying the constraint
\be\bS^2=1. 
\ee
The Lagrangian density is 
\be
\CL={1\over 2 g_0^2}  \partial_\mu {\boldsymbol{S}} \cdot  \partial^\mu {\boldsymbol{S}}, 
\ee
where $g_0$ is the bare coupling constant. The first two coefficients of the beta function are \cite{bzj} 
\be
\beta_0= {1 \over 4 \pi \Delta}, \qquad \beta_1= {1\over 8 \pi^2 \Delta}, 
\ee
where
\be
\label{Delta2} \Delta= {1\over N-2}, 
\ee
and the coefficient $\xi$ defined in (\ref{xi-constant}) is given by 
\be
\xi= \Delta. 
\ee
This theory has a global $O(N)$ symmetry. The conserved currents are given by
\be
J_{\mu}^{ij}=S^{i}\partial_{\mu}S^{j}-S^{j}\partial_{\mu}S^{i},  
\ee
and we will denote by $Q^{ij}$ the corresponding charges. As in \cite{hmn, hn}, 
we take as our charge in (\ref{HQ}) the quantum version of $Q^{12}$. The exact $S$ matrix of the $O(N)$ non-linear sigma model was found in 
\cite{zamo-zamo}. For particles charged under $\mQ^{12}$, it is given by
\be
S(\theta)=-{\Gamma (1+\ri x) \Gamma({1\over 2} +\Delta +\ri x) \Gamma({1\over 2}-\ri x) \Gamma(\Delta-\ri x) \over 
\Gamma (1-\ri x)\Gamma({1\over 2}+\Delta-\ri x) \Gamma({1\over 2}+\ri x) \Gamma(\Delta +\ri x) }, 
\ee
where
\be
\label{xtheta}
x={\theta \over 2 \pi}
\ee
and $\Delta$ is given in (\ref{Delta2}). The mass gap of this model has been known exactly since \cite{hn}
\begin{equation}
\frac{m}{\Lambda}= \left(\frac{8}{\re}\right)^\Delta\frac{1}{\Gamma(1+\Delta)}\,.
\end{equation}

\vskip .2cm

(ii) {\it $\CN=1$ non-linear $O(N)$ sigma model}. The ${\cal N}=1$ supersymmetric version of the non-linear sigma model has 
two fields: the field $\bS$ of the purely bosonic version, satisfying also $\bS^2=1$, and an $N$-uple of Majorana fermions $\boldsymbol{\psi}$ satisfying the constraint
\be
\bS \cdot \boldsymbol{\psi}=0. 
\ee
The Lagrangian density is
\be
\label{susyL}
\CL={1\over 2 g_0^2} \left\{  \partial_\mu {\boldsymbol{S}} \cdot  \partial^\mu {\boldsymbol{S}}+ \ri \,  \overline {\boldsymbol{\psi}} \cdot \slashed{\partial}\boldsymbol{\psi}+ {1\over 4} \left(\overline{\boldsymbol{\psi}} \cdot \boldsymbol{\psi}  \right)^2 \right\}, 
\ee
where we follow the conventions of \cite{witten}. The first two coefficients of the beta function are (see \cite{eh-ssm} and references therein)
\be
\beta_0={1\over 4 \pi \Delta}, \qquad \beta_1=0, 
\ee
where $\Delta$ is given in (\ref{Delta2}). It follows that
\be
\xi=0. 
\ee
The exact $S$ matrix of the supersymmetric $O(N)$ sigma model was obtained in \cite{switten}. 
The precise choice of conserved charge in (\ref{HQ}) is discussed in detail in \cite{eh-ssm}. One eventually obtains an integral 
equation of the form (\ref{chi-ie}) in which the kernel is given by 
\be
\label{rthetasusy}
R(\theta)=\delta(\theta)- K(\theta)= \int_0^\infty {\rd \omega \over \pi} \cos(\omega \theta) 
{\cosh\left( (1-2\Delta) \pi \omega/2 \right) \sinh(\pi \Delta \omega) \over \cosh^2(\pi \omega/2)} \re^{\pi \omega/2}, 
\ee
and $\Delta$ is again as in (\ref{Delta2}).  In \cite{eh-ssm} the mass gap was calculated to be
\begin{equation}
\frac{m}{\Lambda}=2^{2\Delta}\frac{\sin(\pi\Delta)}{\pi\Delta}\,.
\end{equation}

\vskip .2cm

(iii) {\it $SU(N)$ principal chiral field}. Here, the field is a map $\Sigma: \IR^2 \rightarrow SU(N)$, with Lagrangian density
\be
\CL= {1\over  g_0^2} \tr \left(\partial_\mu \Sigma\,  \partial^\mu \Sigma^\dagger\right).
\ee
The first two coefficients of the beta function are \cite{mstone}
\be
\beta_0={1\over 16  \pi \overline \Delta}, \qquad \beta_1={1\over 256 \pi^2 \overline \Delta^2}, 
\ee
where
\be
\label{oDelta}
\overline \Delta ={1\over N}. 
\ee
In this case, 
\be
\xi={1\over 2}. 
\ee
The $S$ matrix of the principal chiral field was obtained in \cite{wiegmann, abdalla}. The choice of conserved charge in (\ref{HQ}) is as in \cite{pcf}, and the relevant $S$ matrix element is 
\be
\label{spcf}
S(\theta)=-{\Gamma^2(1+ \ri x) \Gamma(\overline \Delta- \ri x) \Gamma(1- \overline \Delta -\ri x)  \over 
 \Gamma^2(1- \ri x) \Gamma(\overline \Delta + \ri x) \Gamma(1- \overline \Delta +\ri x)  }, 
 \ee
 where $x$ and $\overline \Delta$ are given in (\ref{xtheta}) and (\ref{oDelta}), respectively. The mass gap is known from \cite{pcf} 
\begin{equation}
\frac{m}{\Lambda}=\sqrt{\frac{8\pi}{\re}}\frac{\sin(\pi\overline \Delta)}{\pi\overline \Delta}\,.
\label{pcf-mg}
\end{equation}
\vskip .2cm

(iv) {\it $SU(N)$ principal chiral field with FKW charges}. The $SU(N)$ principal chiral field can also be explored by using a different set of conserved charges, discussed in \cite{fkw1,fkw2}, 
which despite exciting multiple particles can be more convenient for certain large-$N$ analysis. The kernel in (\ref{chi-ie}) is presented in \cite{fkw2}:
\begin{equation}
R(\theta)=\delta(\theta)- K(\theta)= \frac{1}{2\pi}\int_{-\infty}^\infty \re^{\ri \theta\omega} \frac{\pi\overline \Delta}{2}\frac{\sinh(\pi\overline\Delta|\omega|)}{\cosh(\pi\overline\Delta\omega)-\cos(\pi\overline\Delta)},
\label{kernelfkw}
\end{equation}
where $\overline \Delta$ is given by (\ref{oDelta}).

\vskip .2cm
(v) {\it $O(N)$ Gross--Neveu model}. In the $O(N)$ GN model, the basic fields is an $N$-uple of Majorana fermions $\boldsymbol{\chi}$. 
The Lagrangian density describing the theory is 
\be
\CL= {\ri \over 2} \overline{\boldsymbol{\chi}} \cdot \slashed{\partial} \boldsymbol{\chi}+ {g^2\over 8} \left(\overline{\boldsymbol{\chi}} \cdot \boldsymbol{\chi}  \right)^2, 
\ee
and we follow the conventions of \cite{fnw1,fnw2}. The first two coefficients of the beta function are (see e.g. \cite{gracey})
\be
\beta_0= {1 \over 4 \pi \Delta}, \qquad \beta_1=- {1\over 8 \pi^2 \Delta}, 
\ee
where $\Delta$ is given in (\ref{Delta2}). Therefore, 
\be
\xi=-\Delta. 
\ee
The full $S$ matrix of the GN model was found in \cite{zamo-zamo}. As in \cite{fnw1,fnw2}, we take as our charge in (\ref{HQ}) 
the quantum version of $Q^{12}$, associated to the global $O(N)$ symmetry. The relevant $S$ matrix is then
 \be
 S(\theta)= {\Gamma(1+ \ri x) \Gamma\left({1\over2}-\ri x\right) \Gamma\left(1-\Delta -\ri x\right) \Gamma\left({1\over 2}-\Delta + \ri x \right)
 \over 
 \Gamma(1-\ri x) \Gamma\left({1\over2}+\ri x\right) \Gamma\left(1-\Delta +\ri x\right) \Gamma\left({1\over 2}-\Delta - \ri x \right)}
 \ee
where $x$ and $\Delta$ are again given by (\ref{xtheta}) and (\ref{Delta2}), respectively. The mass gap of this model is was found in \cite{fnw1,fnw2} to be 
\begin{equation}
\frac{m}{\Lambda}=\frac{(2 \re)^\Delta}{\Gamma(1-\Delta)}\,,
\end{equation}
where $\Lambda$ is given by \eqref{Lambda}\footnote{Note that \eqref{Lambda} differs by an overall $2^{-\xi}$ from the convention used in \cite{fnw1,fnw2}.}.

\subsection{General solution of the integral equation}

As it is clear from the discussion above, the perturbative regime of the integrable field theory corresponds to large $h$, which means large $B$ in the 
integral equation. This is a singular limit which is difficult to study analytically. This problem appears already in much simpler models solved by the Bethe ansatz, 
like the Lieb--Liniger \cite{ll} and the Gaudin--Yang \cite{gaudin,yang} models. In a {\it tour-de-force} paper \cite{volin, volin-thesis}, Volin reformulated in a powerful way 
the matching method which was used to study the integral equation 
of the Lieb--Liniger model \cite{hutson, popov}. He applied this method to the non-linear sigma model and he was able to compute analytically the perturbative series 
to large order. 

We have recently generalized Volin's method to solve the long-standing problem of deriving the perturbative series in the Lieb--Liniger and the Gaudin--Yang 
models from the Bethe ansatz \cite{mr-long,mr-ll}. In this paper we will generalize it to the integrable quantum field theories listed above. We will obtain in this way 
analytic results for the solution of the integral equation (\ref{chi-ie}) when $B$ is large, and in particular we will find explicit 
expansions for $\rho$, $e(\rho)$ in power series of $1/B$ and $\log \, B$. In particular, we will streamline 
Volin's method and derive one of its key ingredients directly from the Wiener--Hopf decomposition of the kernel. 

A crucial ingredient in \cite{volin,volin-thesis} (see also \cite{ksv}) is the resolvent of the density of Bethe roots, 
 \be
 \label{resolvent}
 R(\theta)= \int_{-B}^B {\chi(\theta ') \over \theta-\theta'} \rd \theta'. 
 \ee
 This function is analytic in the complex $\theta$-plane but it has a discontinuity in the interval $[-B, B]$, given by 
 \be
 \chi(\theta)=-{1\over 2 \pi \ri} \left( R(\theta+ \ri \epsilon)-R(\theta-\ri \epsilon) \right). 
 \ee
 %
 From its definition we deduce that 
 \be
 \label{Rk}
 R(\theta)= \sum_{k \ge 0} \langle \theta^k\rangle  \theta^{-k-1} , \qquad\langle \theta^k\rangle= \int_{-B}^B \chi(\theta) \theta^k \rd \theta. 
 \ee
It follows from (\ref{erho}) that we can compute the density $\rho$ as a function of $B$ from the residue at infinity of the resolvent. 

The weak coupling regime corresponds to large $B$, so we should study the resolvent 
in a systematic expansion in $1/B$. To do this, we consider the resolvent in two different regimes. The first one is the so-called {\it bulk regime}, in which we take the limit
 \be
 \label{Btheta}
 B\rightarrow \infty, \qquad \theta \rightarrow \infty,
 \ee
 in such a way that 
 \be
 u={\theta \over B}
 \ee
 is fixed. This is therefore appropriate to study $\chi(\theta)$ near $\theta=0$. The second regime is the so-called {\it edge regime}, in which we also have (\ref{Btheta}) but we keep fixed the variable
 \be
 \label{z-var}
 z=2\left(\theta-B\right). 
 \ee
This is therefore appropriate to study $\chi(\theta)$ near the edge of the distribution $\theta=B$. 

\begin{figure}
\center
\includegraphics[height=5cm]{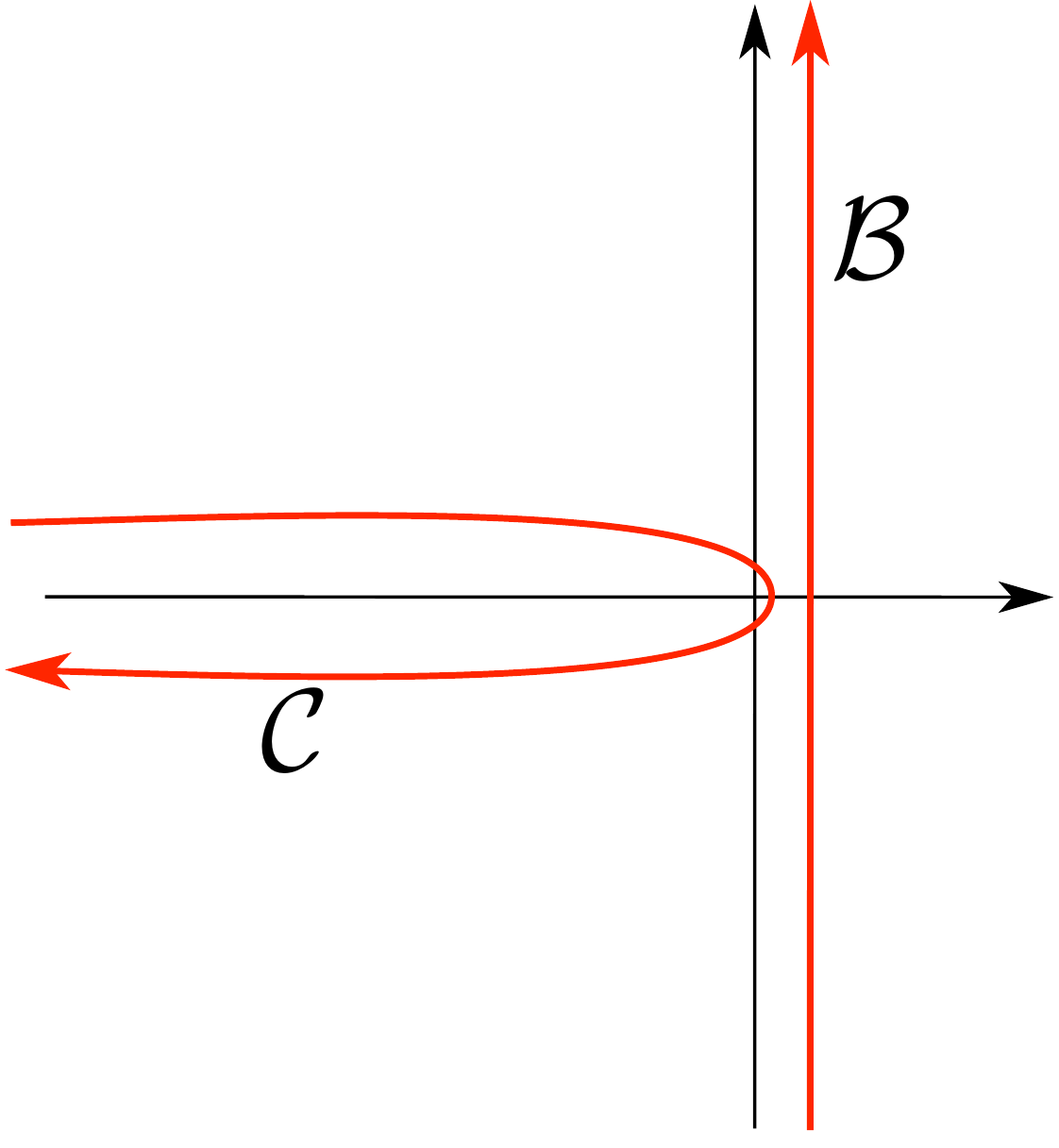}
\caption{The contour $\CC$ in the integral (\ref{e-con}) can be deformed to (minus) the Bromwich contour $\CB$ appearing in the inverse Laplace transform.}
\label{c-contour}
\end{figure}

In order to study the bulk regime, 
we need an appropriate ansatz for $R(\theta)$. We will propose concrete formulae for the relevant models in the next section, following previous studies in \cite{hutson,popov,iw-gy,volin,mr-ll,mr-long}. 
To study the edge regime, we write $R(z)=R(\theta(z))$ as a Laplace transform, 
\be
\label{laplace}
R(z) = \int_0^\infty \hat R(s) \re^{-s z} \rd s. 
\ee
A general property of $\hat R(s)$ is that, as explained in \cite{volin, volin-thesis,mr-long}, it has an expansion as $s\rightarrow \infty$ in integer, negative powers of $s$:
\be
\label{lt-taylor2}
\hat R(s)={\chi_0 \over s}-{\chi_1 \over 2 s^2}+{\chi_2 \over 4 s^3}+\cdots,
\ee
where $\chi_n$, $n \ge 0$, are the coefficients of the Taylor expansion of the density of eigenvalues near $\theta=B$, 
\be
\chi(\theta)= \chi_0 + \chi_1 (\theta-B)+ \chi_2 (\theta-B)^2+ \cdots
\label{taylor}
\ee
The connection between (\ref{lt-taylor2}) and (\ref{taylor}) makes it possible to test 
the solution for $\hat R(s)$ against a numerical study of the distribution $\chi(\theta)$ at the edge. 
One consequence of (\ref{lt-taylor2}) is that, since $\hat R(s)$ decreases as $1/s$ at infinity, one can use the Bromwich inversion formula to write
\be
\label{inv-laplace}
\hat R(s)= \int_{-\ri \infty+\epsilon}^{\ri \infty+ \epsilon} \re^{s z} R(z) {\rd z \over 2 \pi \ri}. 
\ee
The function $\hat R(s)$ can be also used to calculate the energy as a perturbative series in $1/B$ and $\log B$, as pointed out in \cite{volin}. To see this, 
let us neglect exponentially small contributions of the form $\re^{-B}$ in the first equation of (\ref{erho}). Then, we can write
\be
\label{e1}
  \frac{e}{m} \simeq\int_{0}^B\!\! \re^{\theta}\chi(\theta) \frac{\rd\theta}{2\pi}\simeq \re^B\int_{-\infty}^0\re^{z/2}\chi(z)\frac{\rd z}{4\pi}. 
\ee
We have
\be
\label{e-con}
\ba
\int_{-\infty}^0\re^{z/2}\chi(z)\frac{\rd z}{4\pi}&= -{1\over 4 \pi} \int_{-\infty}^0 \re^{z/2} \left( R(z+\ri \epsilon) - R(z-\ri \epsilon) \right) {\rd z\over 2 \pi \ri}\\
& ={1\over 4 \pi} \int_{{\cal C}}\re^{z/2} R(z)  {\rd z\over 2 \pi \ri}, 
\ea
\ee
where the contour ${\cal C}$ is shown in \figref{c-contour}. It can be deformed to (minus) the Bromwich contour, and we obtain at the end of the day
\be
\label{eR}
{e\over m} = {\re^B \over 4 \pi} \hat R(1/2), 
\ee
up to exponentially small corrections at large $B$. 

It follows from the above considerations that, if one knows the resolvent $R(\theta)$ and its inverse 
Laplace transform $\hat R(s)$, it is possible to calculate the functions $\rho$, $e$ as a function 
of $B$. To determine $R(\theta)$, there are two possible routes. In the first one, followed in \cite{volin}, 
one writes the kernel $K(\theta)$ in the form
\be
\label{ko}
K(\theta)= \mO {1\over \theta}, 
\ee
where $\mO$ is a difference operator. The integral equation can then be written as a difference and discontinuity equation for the resolvent, 
 \be
 -{1\over 2 \pi \ri} \left( R(\theta+ \ri \epsilon)-R(\theta-\ri \epsilon) \right)- \mO R(\theta)= m \cosh\, \theta. 
 \ee
This can be used in the edge regime to determine the analytic structure of $\hat R(s)$, which makes it possible to derive its functional form. However, in 
\cite{mr-long} it was shown, based on observations in \cite{fnw1,volin}, that one can find $\hat R(s)$ directly from the Wiener--Hopf decomposition of 
the kernel $K(\theta)$. Let us present the argument in \cite{mr-long}, adapted to the setting of this paper. 

We first use the variable (\ref{z-var}) to inspect the edge limit of (\ref{chi-ie}). It is useful to define $\chi(z)=\chi(\theta(z))$ and
\begin{align}
F_-(\omega)&=\int_{-\infty}^{\infty} \re^{\ri\omega z} \chi(z)(1-\Theta(z)) \rd z=\int_{-\infty}^0 \re^{\ri\omega z} \chi(z) \rd z\,, \label{eq_fminus}\\
\widehat{K}(\omega)&=\int_{-\infty}^{\infty} \re^{\ri\omega \theta}K(\theta)\rd \theta, 
\end{align}
where $\Theta(z)$ is the Heaviside function. Then, 
\begin{equation}
\int_{-\infty}^\infty \re^{-\ri\omega z}(1-\widehat{K}(2\omega))F_-(\omega)\rd \omega= \frac{m \re^B}{2} \re^{z/2}+\mathcal{O}(\re^{-B}),\quad  z<0. 
\end{equation}
By extending the above equation to all real $z$, and ignoring exponentially small terms when $B$ is large, we obtain
\begin{equation}
\int_{-\infty}^\infty \re^{-\ri\omega z}(1-\widehat{K}(2\omega))F_-(\omega)\rd \omega= (1-\Theta(z))\frac{m \re^B}{2} \re^{z/2}+\Theta(z)\xi(z), \label{eq_preF}
\end{equation}
where $\xi(z)$ is an unknown function. Let us then define one last Fourier transform so that we can transform the full equation:
\begin{equation}
X_+(\omega)=\int_{-\infty}^\infty \re^{\ri\omega z} \xi(z)\Theta(z) \rd z=\int^{\infty}_0 \re^{\ri\omega z} \xi(z) \rd z\,.
\end{equation}
The subscripts $\pm$ denote that something is analytic in the upper/lower half complex plane (including the real axis 
but possibly excluding the origin). We also introduce the Wiener-Hopf decomposition of the kernel
\be 
1-\widehat{K}(\omega)=\frac{1}{G_+(\omega)G_-(\omega)},
\ee
where $G_+(\omega)=G_-(-\omega)$ if $\widehat{K}(\omega)$ is even (as it happens in all the cases we will consider). This 
decomposition can almost always be done provided $1-\widehat{K}(\omega)$ is well defined along the real axis (see e.g. \cite{fnw1}), 
though some care might be necessary at $\omega=0$. With the above definitions, we take the Fourier transform of (\ref{eq_preF}) and obtain
\begin{equation}
\frac{F_-(\omega)}{G_+(2\omega)G_-(2\omega)} = \frac{ m \re^B }{2\ri}\frac{1}{\omega-\ri/2} +  X_+(\omega)\,.
\label{eq_WH0}
\end{equation}
We can rewrite (\ref{eq_WH0}) as
\begin{align}
\frac{F_-(\omega)}{G_-(2\omega)} - \frac{m\re^B}{2\ri}\frac{G_+(\ri)}{\omega-\ri/2}&= \frac{m\re^B}{2\ri}\frac{G_+(2\omega)-G_+(\ri)}{\omega-\ri/2}+G_+(2\omega)X_+(\omega)= C(\omega)\,.
\label{eq_WH}
\end{align}
From \eqref{eq_WH} it follows that $C(\omega)$ must be analytic in both the upper and lower half complex planes, reducing it to an entire function. However, in their respective half planes (including the real axis), $F_-(\infty)=0$, $X_\pm(\infty)=0$ and $G_\pm(\infty)= \text{constant}$ (where the first two come from their 
definitions and \eqref{eq_preF}, while the latter can be checked explicitly). Both sides are thus bound at infinity, and by using Liouville's Theorem 
we find that $C(\omega)=0$. We conclude that, at leading order in the edge limit,
\begin{equation}
F_-(\omega)= m \re^B \frac{G_+(\ri)}{2}\, \frac{G_+(-2\omega)}{\ri\omega+1/2}\,.
\label{eq_WH_basic}
\end{equation}

We can now relate this function to the inverse Laplace transform of the resolvent (\ref{inv-laplace}). 
We consider $\hat{R}(\ri s)$ for ${\rm Im}(s)<0$ and bend the Bromwich contour around the negative real axis, 
without crossing it, as in \figref{c-contour}. We obtain
\begin{align*}
\hat{R}(\ri s)&=\int_{-\infty+\ri\epsilon}^{0+\ri\epsilon}\frac{\re^{\ri s z}  }{2\pi \ri}R(z) \rd z-\int_{-\infty-\ri\epsilon}^{0-\ri\epsilon}\frac{\re^{\ri s z}  }{2\pi \ri}R(z) \rd z=\int_{-\infty}^{0}\e^{\ri s z} \left(\frac{R(z-\ri\epsilon)-R(z+\ri\epsilon)}{2\pi \ri}\right)\rd z\\
&=\int_{-\infty}^{0}\re^{\ri s z} \chi(z) \rd z = F_-(s) + \mathcal{O}(\re^{-B}).
\end{align*}
This is the leading solution in the strict large $B$ limit, but one has additional corrections as a 
power series in $1/B$. Taking into account constraints on the allowed poles and behaviour at infinity of $\hat{R}(s)$, we can write 
\begin{equation}
\label{edge-sol}
\hat{R}(s)=m \, \re^B A \Phi(s)\left(\frac{1}{s+\frac{1}{2}}+Q(s)\right),
\end{equation}
where
\be
\label{exp-sol}
\Phi(s)= G_+(2 \ri s), \qquad A= {G_+(\ri) \over 2}, 
\ee
and $Q(s)$ is of the form
\be
  Q(s)=\frac 1{Bs} \sum_{n,m=0}^\infty { Q_{n,m}(\log B) \over B^{m+n} s^{n}}.
\ee
The result (\ref{edge-sol}) is of the form obtained by Volin in \cite{volin} in the case of the non-linear sigma model. In our derivation, 
the function $\Phi(s)$ is determined simply by the Wiener--Hopf decomposition of the kernel, therefore it can be written immediately in a large number of cases. 
The coefficients $Q_{n,m}$ appearing in this expansion are not fixed by the Wiener--Hopf decomposition, but as shown in \cite{hutson,popov, volin}, they can be calculated 
recursively by comparing the edge solution (\ref{edge-sol}) to the ansatz for $R(\theta)$ in the bulk regime, see \cite{volin, mr-long} for details on the matching procedure. 

\subsection{Solving the bosonic models}

As we mentioned before, we will refer to the non-linear sigma model, its supersymmetric extension, and the PCF, as ``bosonic" models. In the bosonic models the perturbative expansion of the free energy has the structure
\be
\mathcal{F}(h)-\mathcal{F}(0)=-h^2 \kappa_0 \left\{ \frac{1}{\overline{g}^2}+\beta_0 \kappa_1+\beta_0^2\kappa_2 \gbar+\mathcal{O}\left(\gbar^3\right) \right\}, 
\end{equation}
where $\kappa_i$, $i=0,1,2$ are calculable constants and $\bar g$ is the RG-invariant coupling defined in 
(\ref{RG-evol}). In order to write down the perturbative series, it is useful to follow \cite{bbbkp} 
and introduce an intermediate 
coupling $\widetilde \alpha$ defined as\footnote{In \cite{bbbkp} this coupling is denoted by $\alpha$, but we 
reserve this simpler notation for our final choice of coupling (\ref{def_agen}), 
which plays a more important r\^ole in the solution of the integral equation.}
\begin{equation}
\label{aldef}
\frac{1}{\widetilde \alpha}+\xi \log \widetilde \alpha=\log\left( \frac{h}{\Lambda}\right),
\end{equation}
where $\xi$ is given in (\ref{xi-constant}). Similar schemes in QCD have been considered in \cite{bly, jamin-mira}. We have, at leading order, 
\be
\widetilde \alpha\approx 2\beta_0 \gbar^2, \qquad \rho \approx { \kappa_0 h \beta_0 \over \widetilde \alpha}. 
\ee
This suggests to introduce yet another coupling constant $\alpha$ through the relation 
\begin{equation}
\frac{1}{\alpha}+(\xi-1)\log\alpha=\log\left(\frac{\rho}{ 2\mathfrak{c} \beta_0\Lambda}\right),
\label{def_agen}
\end{equation}
where $\mathfrak{c}$ is a model-dependent constant. As we will see, the quotient $e/\rho^2$ can be expressed as a 
formal power series in $\alpha$, without logarithms. A wise choice of $\mathfrak{c}$ simplifies the resulting expressions (removing e.g. terms involving 
$\log(2)$). 

Let us now consider the solution of the integral equation for the bosonic models. 
We have, as noted in \cite{pcf} in the case of the PCF, 
\begin{equation}
\label{Gexp}
G_+(2\ri s)=\frac{k}{\sqrt{s}} \re^{\eta s} \re^{a s \log(s)} \phi(s)=\frac{k}{\sqrt{s}}  \re^{a s \log(s)} (1-b s+\cdots),
\end{equation}
where $k$, $\eta$, $a$ are constants. $\phi(s)$ is typically a rational function of $\Gamma$-functions, and there are 
no $\log(s)$ on the r.h.s. before expanding the last exponential. An interesting observation 
is that $\eta$ does not affect the final result for the energy expansion. In \cite{pcf} it is noted that we must have $a=1-2\xi$. It is convenient to normalize
\begin{equation}
\Phi(s) = \frac{G_+(2\ri s)}{k\sqrt{\pi}}=\frac{1}{\sqrt{\pi s}}+\cdots
\end{equation}
We re-define $A$ to keep the form of $\hat{R}(s)$ unchanged:
\begin{equation}
\hat{R}(s)=m \, \re^B \, A \Phi(s)\left(\frac{1}{s+\frac{1}{2}}+Q(s)\right),\quad A= \frac{k \sqrt{\pi}}{2}G_+(\ri)=\frac{k^2\pi}{2}\Phi(1/2).
\label{edge}
\end{equation}
From this formula we extract directly, by using (\ref{eR}), 
\begin{equation}
e= 
\frac{m^2 \re^{2B} A}{4\pi}\Phi(1/2)\left(1+\sum_{m=0}^\infty\frac{1}{B^m}\sum_{s=0}^{m-1}
2^{s+1}Q_{s,n-1-s}\right)\equiv m^2 \e^{2B} \frac{A^2}{2\pi^2k^2}\tilde{e}.
\label{epst}
\end{equation}
In order to fix the remaining coefficients, we need an ansatz for the resolvent in the bulk. For the bosonic models we 
can use Volin's ansatz in \cite{volin}, 
\begin{equation}
R(\theta)=\sum_{m=0}^\infty\sum_{n=0}^\infty\sum_{k=0}^{m+n}2A\sqrt{B} c_{n,m,k}\frac{(\theta/B)^{e(k)}}{B^{m-n}(\theta^2-B^2)^{n+1/2}}\left[ \log\left(\frac{\theta-B}{\theta+B}\right) \right]^k, 
\label{bulk}
\end{equation}
where $e(k)=0$ if $k$ is even and $1$ if $k$ is odd. Note that with the above conventions we have $c_{0,0,0}=1$. From (\ref{erho}) and (\ref{Rk}) we obtain 
\begin{equation}
\rho= \frac{2 m \re^B \sqrt{B}A}{2\pi}\left(1+\sum_{m=1}^\infty\frac{c_{0,m,0}-2 c_{0,m,1}}{B^m}\right)\equiv \frac{m \re^B\sqrt{B}A}{\pi}\tilde{\rho},
\label{rhot}
\end{equation}
leading to the convenient normalization
\begin{equation}
\frac{e}{\rho^2}=\frac{1}{2 k^2}\frac{\tilde{e}}{B \tilde{\rho}^2}.
\end{equation}
Let us now list the results for the different models. 

\vskip .2cm

(i) {\it Non-linear $O(N)$ sigma model}. In this case, the kernel has the Fourier transform
\cite{hn}
\be
1-\widehat K(\omega)=\frac{1-\re^{-2\pi\Delta|\omega|}}{1+\re^{-\pi|\omega|}}, 
\ee
where $\Delta$ is given in (\ref{Delta2}). 
Its Wiener--Hopf decomposition is determined by
\be
G_+(\omega)=\frac{\re^{-{\ri \omega \over 2} ((1-2\Delta)(\log(-\ri \omega/2)-1)-2\Delta\log(2\Delta))}}{\sqrt{-\ri \Delta \omega}}\frac{\Gamma(1-\ri \Delta \omega)}{\Gamma(1/2-\ri \omega/2)},
\ee
which gives
\begin{align}
\Phi(s)&=\frac{\re^{s((1-2\Delta)(\log(s)-1)-2\Delta\log(2\Delta))}}{\sqrt{ s}}\frac{\Gamma(1+2\Delta s)}{\Gamma(1/2+s)},\\
A&=\frac{\re^{-\frac{1}{2}+\Delta}\Delta^{-\Delta}\Gamma(\Delta)}{4 }, \qquad k=\frac{1}{\sqrt{2\pi\Delta}}, 
\end{align}
and we recover the results of Volin for this model \cite{volin}. The coupling constant $\alpha$ is defined by (\ref{def_agen}) with $\mathfrak{c}=1$. The 
perturbative series reads
\begin{equation}
\label{on-ps}
\ba
\frac{\tilde{e}}{B\tilde{\rho}^2}&=\alpha +\frac{\alpha ^2}{2}+\frac{\alpha ^3 \Delta }{2}+\frac{1}{32} \alpha ^4 \Delta  \left(-8 \Delta ^2 (3 \zeta (3)+1)+14 \Delta  (3 \zeta (3)+2)-21 \zeta (3)+8\right)\\
&+\frac{1}{96} \alpha ^5 \Delta  \left(-24 \Delta ^3 (19 \zeta (3)+1)+\Delta ^2 (918 \zeta (3)+60)-7 \Delta  (87 \zeta (3)-20)+3 (35 \zeta (3)+8)\right)\\
&+\frac{1}{6144}\alpha ^6 \Delta  \left(-96 \Delta ^4 (1024 \zeta (3)+405 \zeta (5)+10)+24 \Delta ^3 (8544 \zeta (3)+4185 \zeta (5)+4)\right.\\
&\left.\quad-8 \Delta ^2 (19236 \zeta (3)+12555 \zeta (5)-2200)+12 \Delta  (3878 \zeta (3)+93 (45 \zeta (5)+16))\right.\\
&\left.\quad-9 (980 \zeta (3)+1395 \zeta (5)-256)\right)\\
&+ \CO(\alpha^7).
\ea
\end{equation}
We have calculated analytically the first $44$ terms of this expansion. 

\vskip .2cm
(ii) {\it $\CN=1$ non-linear $O(N)$ sigma model}. The Fourier transform of the kernel can be obtained immediately from 
(\ref{rthetasusy}):
\be
1-\widehat K(\omega)=\frac{\cosh((1-2\Delta)\pi|\omega|/2)\sinh(\pi\Delta|\omega|)}{\cosh(\pi|\omega|/2)^2}\re^{\pi|\omega|/2}, 
\ee
where $\Delta$ is given in (\ref{Delta2}). The Wiener--Hopf decomposition was obtained in \cite{eh-ssm}
\begin{align}
\Phi (s)&=\frac{\re^{s(-2(1-\log(s))+(1-2\Delta)(1-\log((1-2\Delta)s)))+2\Delta(1-\log(2\Delta s)))}}{\sqrt{s}}\\
& \times \frac{\Gamma\left(\frac{1}{2}+(1-2\Delta)s\right)\Gamma(1+2\Delta s)}{\Gamma\left(\frac{1}{2}+s\right)^2}, 
\end{align}
and one finds, 
\be
A=\frac{2^{-2-\Delta}\re^{-1/2}\pi(1-2\Delta)^{\Delta-1/2}\Delta^{-\Delta}}{\sin(\pi\Delta)},  \qquad k=\frac{1}{\sqrt{2\pi\Delta}}.
\ee
Using the same coupling constant as in the non-linear sigma model, i.e. (\ref{def_agen}) with $\mathfrak{c}=1$, we obtain the perturbative expansion
\begin{equation}
\label{susyon-ps}
\ba
\frac{\tilde{e}}{B\tilde{\rho}^2}&=\alpha +\frac{\alpha ^2}{2}-\frac{3}{32} \alpha ^4 \left(\Delta  \left(8 \Delta ^2-14 \Delta +7\right) \zeta (3)\right)+\frac{5}{32} \alpha ^5 \Delta  \left(8 \Delta ^2-14 \Delta +7\right) \zeta (3)\\
&-\frac{15}{2048} \alpha ^6 \Delta  \left(864 \Delta ^4 \zeta (5)-2232 \Delta ^3 \zeta (5)+8 \Delta ^2 (28 \zeta (3)+279 \zeta (5))\right.\\
&\left.\quad-4 \Delta  (98 \zeta (3)+279 \zeta (5))+196 \zeta (3)+279 \zeta (5)\right)\\
&+\frac{3}{1024} \alpha ^7 \Delta  \left(3648 \Delta ^5 \zeta (3)^2-672 \Delta ^4 \left(19 \zeta (3)^2-13 \zeta (5)\right)+28 \Delta ^3 \left(627 \zeta (3)^2-806 \zeta (5)\right))\right.\\
&\left.\quad-28 \Delta ^2 \left(399 \zeta (3)^2-24 \zeta (3)-806 \zeta (5)\right)+7 \Delta  \left(399 \zeta (3)^2-168 \zeta (3)-1612 \zeta (5)\right))\right.\\
&\left.\quad+7 (84 \zeta (3)+403 \zeta (5))\right)+\CO(\alpha^8).\\
\ea
\end{equation}
We have calculated analytically the first $42$ terms in this expansion. 
\vskip .2cm
(iii) {\it $SU(N)$ principal chiral field}. From the S-matrix (\ref{spcf}) we can extract
\be
1-\widehat K(\omega)=\frac{2\sinh(\pi\overline{\Delta} |\omega|)\sinh((1-\pi\overline{\Delta})|\omega|)}{\sinh(\pi|\omega|)},
\ee
where $\overline \Delta$ is defined in (\ref{oDelta}). This leads to
\be
\ba
\Phi(s)&=\frac{\re^{-2(\overline{\Delta}\log\overline{\Delta}+(1-\overline{\Delta})\log(1-\overline{\Delta}))s}}{\sqrt{\pi s}}\frac{\Gamma(2\overline{\Delta} s+1)\Gamma(2(1-\overline{\Delta})s+1)}{\Gamma(2 s+1)},\\
A&=\frac{\sqrt{\pi}}{4\sqrt{2}} {\overline{\Delta}^{-\overline{\Delta}}(1-\overline{\Delta})^{\overline{\Delta}-1} \over \sin(\pi\overline{\Delta})},\quad k=\frac{1}{2\sqrt{\pi(1-\overline{\Delta}) \overline{\Delta} }}.
\ea
\ee
Due to the absence of a $\log(s)$ in $\Phi(s)$, all coefficients $c_{n,m,k}$ in (\ref{bulk}) with $k\neq0$ vanish. The coupling $\alpha$ is defined by (\ref{def_agen}) with $\mathfrak{c}=4$, and we obtain:
\begin{equation}
\ba
\frac{\tilde{e}}{B\tilde{\rho}^2}&=\alpha +\frac{\alpha ^2}{2}+\frac{\alpha ^3}{4}+\frac{1}{16} \alpha ^4 \left(6 \overline{\Delta} ^2 \zeta (3)-6 \overline{\Delta}  \zeta (3)+5\right)+\frac{1}{96} \alpha ^5 \left(54 \overline{\Delta} ^2 \zeta (3)-54 \overline{\Delta}  \zeta (3)+53\right)\\
&+\frac{1}{384} \alpha ^6 \left(405 \overline{\Delta} ^4 \zeta (5)-810 \overline{\Delta} ^3 \zeta (5)+81 \overline{\Delta} ^2 (7 \zeta (3)+10 \zeta (5))-81 \overline{\Delta}  (7 \zeta (3)+5 \zeta (5))+487\right)\\
&+\frac{1}{3840}\alpha ^7 \left(135 \overline{\Delta} ^4 \left(76 \zeta (3)^2+75 \zeta (5)\right)-270 \overline{\Delta} ^3 \left(76 \zeta (3)^2+75 \zeta (5)\right)\right.\\
&\left.\quad+45 \overline{\Delta} ^2 \left(228 \zeta (3)^2+391 \zeta (3)+450 \zeta (5)\right)-45 \overline{\Delta}  (391 \zeta (3)+225 \zeta (5))+13804\right)\\
&+\CO\left(\alpha ^8\right).
\ea
\label{pcf-ps}
\end{equation}
We have calculated analytically the first $54$ terms of this expansion. 
\vskip .2cm

(iv) {\it $SU(N)$ principal chiral field with FKW charges}. 
From (\ref{kernelfkw}) we retrieve
\begin{equation}
\ba
\Phi(s)&=\frac{ \overline\Delta \sqrt{\pi }}{\sin \left(\frac{\pi  \overline\Delta }{2}\right)\sqrt{s}}  \frac{\re^{-2 \overline\Delta \log(2) s}\left(s+\frac{1}{2}\right) \Gamma (2 s \overline\Delta +1)}{ \Gamma \left(s \overline\Delta -\frac{\overline\Delta }{2}+1\right) \Gamma \left(s \overline\Delta +\frac{\overline\Delta }{2}+1\right)},\\
 A&=\frac{2^{\frac{1}{2}-\overline\Delta } \sin \left(\frac{\pi  \overline\Delta }{2}\right)}{\overline\Delta \sqrt{\pi } },\qquad k=\frac{\sqrt{2} \sin \left(\frac{\pi  \overline\Delta }{2}\right)}{\pi  \overline\Delta  }.
\ea
\end{equation}

Due to the existence of multiple particles we must change the overall factors in the definitions of the observables, though we keep the definitions of $\tilde{e}$ and $\tilde{\rho}$ from (\ref{epst}) and (\ref{rhot}) respectively,
\be
\ba
\rho_{\rm FKW}&=\frac{1}{4}\int_{-B}^B \chi(\theta)\rd\theta= \frac{m\re^B A}{2}\sqrt{B}\tilde{\rho}.\\
e_{\rm FKW}&=\frac{m}{8\sin^2(\pi\overline\Delta)}\int_{-B}^B \chi(\theta)\cosh\theta\rd\theta  = \frac{m^2 \re^{2B}}{8\sin^2(\pi\overline\Delta)}\frac{A \Phi(1/2)}{2}\tilde{e}.
\label{rhoefkw}
\ea
\ee
As in the standard PCF case discussed above, only the $c_{n,m,0}$ coefficients are non-zero. This is similar to the analysis done in \cite{kazakov19}. From now on, we will remove the subscript ${\rm FKW}$ in $e$, $\rho$ and work exclusively with the quantities defined in (\ref{rhoefkw}).

We introduce the coupling constant defined in (\ref{def_agen}) and we choose $\mathfrak{c}$ to be
\begin{equation}
\mathfrak{c}= \frac{16}{\overline\Delta\sqrt{\re}}\sin \left(\frac{\pi  \overline\Delta }{2}\right) \sin (\pi  \overline\Delta ) \re^{\frac{\overline\Delta}{2}   \left(\psi^{(0)}\left(1+\frac{\overline\Delta }{2}\right)+\psi^{(0)}\left(1-\frac{\overline\Delta }{2}\right)+2\gamma_E \right)},
\end{equation}
 where $\gamma_E$ is the Euler-Mascheroni constant and $\psi^{(m)}$ is the polygamma function. With this choice of $\alpha$ we find
 \begin{equation}
 \ba
 \frac{\tilde{e}}{B \tilde{\rho}^2}&= \alpha +\frac{\alpha ^3}{4}+\frac{1}{8} \alpha ^4 \left(\overline\Delta ^3 Z_{\overline\Delta}(3)-1\right)+\frac{1}{48} \alpha ^5 \left(20-3 \overline\Delta ^3 Z_{\overline\Delta}(3)\right)
 \\&+\frac{1}{384} \alpha ^6 \left(81 \overline\Delta ^5 Z_{\overline\Delta}(5)+177 \overline\Delta ^3 Z_{\overline\Delta}(3)-110\right)
 \\&+\frac{\alpha ^7 \left(-405 \overline\Delta ^5 Z_{\overline\Delta}(5)+15 \overline\Delta ^3 Z_{\overline\Delta}(3) \left(76 \overline\Delta ^3 Z_{\overline\Delta}(3)-83\right)+6646\right)}{3840}
 \\&+\frac{\alpha ^8 \left(16875 \overline\Delta ^7 Z_{\overline\Delta}(7)-2700 \overline\Delta ^6 Z_{\overline\Delta}(3){}^2+27270 \overline\Delta ^5 Z_{\overline\Delta}(5)+48935 \overline\Delta ^3 Z_{\overline\Delta}(3)-18332\right)}{15360}
 \\&+ \frac{\alpha ^9}{645120} \left(2313360 \overline\Delta ^8 Z_{\overline\Delta}(3) Z_{\overline\Delta}(5)-354375 \overline\Delta ^7 Z_{\overline\Delta}(7)+2208780 \overline\Delta ^6 Z_{\overline\Delta}(3){}^2\right.+
\\&\quad+ \left. -708750 \overline\Delta ^5 Z_{\overline\Delta}(5)-1429155 \overline\Delta ^3 Z_{\overline\Delta}(3)+10127668\right)\\
 &+\CO\left(\alpha ^{10}\right).
 \ea
 \label{efkw}
 \end{equation}
For compactness we have introduced the auxiliary function
\begin{equation}
Z_{\overline\Delta}(n)=\zeta(n)+\frac{(-1)^{n+1}}{2^n\Gamma(n)}\left(\psi^{(n-1)}\left(1+\frac{\overline\Delta}{2}\right)+\psi^{(n-1)}\left(1-\frac{\overline\Delta}{2}\right)\right).
\end{equation}
We have calculated the first $50$ terms of the expansion (\ref{efkw}). 

It is instructive to test the above expansion against one-loop perturbation theory (in the other models compared in this paper, this comparison has been done in \cite{hmn,hn,pcf,fnw1,fnw2,h-pcf,eh-ssm} to derive the mass gap). 
In order to do this, we have to compute the free energy as a function of the external field $h$. 
We find 
\be
h = 2\sin^2(\pi \overline{\Delta}) \frac{\rd e(\rho)}{\rd\rho},\qquad \delta\mathcal{F}(h)=\mathcal{F}(h)-\mathcal{F}(0)= e-\frac{h\rho}{ 2\sin^2(\pi \overline{\Delta})}. 
\end{equation}
We have to express first $\alpha$ in terms of $h/m$:
\begin{equation}
\frac{1}{\alpha}=\log\frac{h}{m}+\frac{1}{2}\log\log\frac{h}{m}+ \log \left(\frac{\sqrt{2} \sin \left(\frac{\pi  \overline{\Delta} }{2}\right)}{\sqrt{\pi}  \overline{\Delta} }\right)-\frac{\overline{\Delta}}{2}\eta_{\overline{\Delta}} + \CO\left(\frac{1}{\log\frac{h}{m}}\right),
\end{equation}
where
\be
\eta_{\overline{\Delta}}=\psi\left(1+\frac{\overline{\Delta} }{2}\right)+\psi\left(1-\frac{\overline{\Delta} }{2}\right)+2 \gamma_E .
\ee
By using all these results, one eventually finds, 
\begin{equation}
\ba
f(h)=&-\frac{16\pi\overline{\Delta}^2 \delta\mathcal{F}(h)}{h^2}\\
=&\frac{1}{\cos\left(\frac{\pi\overline{\Delta}}{2}\right)^2}\left(\log \frac{h}{m}+\frac{1}{2}\log \log \frac{h}{m}+ \log \left(\frac{\sqrt{2} \sin \left(\frac{\pi  \overline{\Delta} }{2}\right)}{\sqrt{\pi}  \overline{\Delta} }\right)-\frac{\overline{\Delta}}{2}\eta_{\overline{\Delta}}+ \CO\left(\frac{1}{\log\frac{h}{m}}\right)\right).
\ea
\label{pertBA}
\end{equation}
Let us now compare this result to the one-loop expression for the free energy. After using the relationship \eqref{pcf-mg} found in \cite{pcf} between $m$ 
and the dynamically generated scale $\Lambda
$, 
%
%
we obtain 
\begin{equation}
\ba
f_{\text{1-loop}}(h)=& 2\overline{\Delta}\left(\sum _{k=1}^{N} q_k^2 \right)\left( \log \frac{h}{m}+\frac{1}{2}\log \log \frac{h}{m}+\log \left(\frac{\sqrt{2} \sin (\pi  \overline{\Delta} )}{\sqrt{\pi\re}  \overline{\Delta} }\right)\right)\\& + 2\overline{\Delta}^2\sum _{1\le i<j\le N} (q_j-q_i)^2 \left(\log |q_j-q_i|-\frac{1}{2}\right) +\CO\left(\frac{1}{\log\frac{h}{m}}\right),
\ea
\label{balogf}
\end{equation}
where the charges $q_i$ are given by \cite{fkw1}
\begin{equation}
q_k=\frac{\cos\left(\pi\overline{\Delta}\left(k-1/2\right)\right)}{\cos(\pi\overline{\Delta}/2)}. 
\end{equation}
Since
\be
\sum _{k=1}^{N} \cos^2\left(\pi\overline{\Delta}\left(k-1/2\right)\right)={1\over 2 \overline{\Delta}},
\ee
the $h$-dependent term of (\ref{pertBA}) matches the one in (\ref{balogf}). In order for the $h$-independent terms to match, the following equality must hold
\be
\label{st-eq}
\sum _{1\le i<j\le N} (q_j-q_i)^2 \left(\log |q_j-q_i|-\frac{1}{2}\right) = {1\over 2 \left( \overline{\Delta} \cos(\pi \overline{\Delta}/2) \right)^2} 
\left\{ \log \left( { \sqrt{\re} \over 2 \cos(\pi \overline{\Delta}/2)} \right)- 
{\overline{\Delta}\over 2} \eta_{\overline{\Delta}} \right\}. 
\ee
We do not have an analytic proof of this identity, but we have checked it for many values of $N$. 

In \cite{fkw1} and more recently in \cite{kazakov19}, the PCF with FKW charges has been studied in the large $N$ expansion. Although we work at finite $N$, it is straightforward to expand our results in series in $1/N$, and 
we find for example 
\begin{equation}
\ba
f(h)&=\log \frac{h}{m}+\frac{1}{2}\log \log \frac{h}{m}+\frac{1}{2}\log \frac{\pi }{2}+\\
&+\overline{\Delta} ^2\frac{\pi^2 }{24} \left(6 \log \frac{h}{m}+3\log \log \frac{h}{m}+3 \log \frac{\pi}{2}-1\right)+\overline{\Delta} ^3\frac{ \zeta (3)}{4 }+\CO\left(\overline{\Delta} ^4\right),
\ea
\end{equation}
which agrees with the results in \cite{fkw1,kazakov19}.

\subsection{Solving the Gross--Neveu model}

In the case of the GN model, the free energy $\CF(h)$ was calculated in perturbation theory in \cite{fnw1}, and it reads, 
 \be
\CF(h) -\CF(0)= -{h^2\over 2 \pi} \left\{ 1-{\overline g^2  \over 2 \pi}+{K_N \over 4 \pi^2} \overline g^4 + \CO(\overline g^6) \right\}, 
 \ee
 where  
 \be
 K_N= {\log(2)-1 \over \Delta}+{1\over 2}, 
 \ee
and  $\overline g$ is the RG invariant coupling constant defined by (\ref{RG-evol}). One can introduce an intermediate coupling exactly as in (\ref{aldef}), 
but in order to simplify the perturbative expansion it is better to use the analogue of (\ref{def_agen}), which in the GN case it is given by
\be
\label{alpha-GN}
\frac{1}{\alpha}+\xi  \log\alpha=\log\left(\frac{2\pi\rho}{\Lambda}\right). 
\end{equation}

Let us now address the solution of the integral equation. The Fourier transform of the kernel gives
\begin{equation}
1-\widehat K(\omega)=\frac{1+\re^{-2\pi  \tilde{\Delta}|\omega|}}{1+\re^{-\pi  |\omega|}}, 
\end{equation}
where
\be
 \tilde{\Delta}=\frac{1}{2}-\Delta 
 \ee
and $\Delta$ is defined in (\ref{Delta2}). The Wiener--Hopf decomposition is determined by
\begin{equation}
G_+(\omega)=\re^{\frac{\ri\omega}{2}(2\tilde{\Delta}(\log(-\tilde{\Delta}\ri\omega)-1)-(\log(-\ri\omega/2)-1))}\frac{\Gamma\left(\frac{1}{2}
-\ri\tilde{\Delta}\omega\right)}{\Gamma\left(\frac{1}{2}-\frac{\ri\omega}{2}\right)}.
\end{equation}
From (\ref{exp-sol}) we find
\be
\hat{R}(s)= m \, \re^B A \Phi(s)\left( \frac{1}{s+1/2}+\frac{1}{B s}\sum_{m=0}^\infty\sum_{n=0}^m\frac{Q_{n,m-n}(\log(B)) }{B^m s^n}\right)\,,
\label{R_GN}
\ee
where
\be
\ba
\Phi(s)&=\re^{(1-2 \tilde{\Delta} ) s \log \left(\frac{s}{e}\right)-2 \tilde{\Delta} s \log (2 \tilde{\Delta})} \frac{\Gamma \left(2 \tilde{\Delta} s +\frac{1}{2}\right)}{\Gamma \left(s+\frac{1}{2}\right)},\\
A&=\frac{\re^B}{2^{3/2}} \frac{\re^{\tilde{\Delta}-1/2}}{\tilde{\Delta}^{\tilde{\Delta}}}\Gamma(1-\Delta).
\ea
\ee
We also need an ansatz for the resolvent in the bulk. In the GN case, we have
\begin{equation}
R(\theta)=\sum_{m=1}^\infty\frac{A}{B^m}\sum_{n=1}^m\sum_{k=0}^m c_{n,m-n,k}\frac{(\theta/B)^{e(k)}}{(\theta^2/B^2-1)^n}\log\left(\frac{\theta-B}{\theta+B}\right)^k,
\label{bulkGN}
\end{equation}
where $e(k)=0$ if $k$ is odd and $1$ if $k$ is even. This has a different structure from the bosonic case, but not surprisingly it is similar to the bulk ansatz used for the Gaudin--Yang model in \cite{mr-long, mr-ll}. 
From the bulk and the edge ansatz, we obtain the following expressions for $\rho$ and $e$:
\begin{align}
\rho&=\frac{m \, \re^{B}A} {2\pi}\sum_m \frac{c_{1,m,0}}{B^m}\equiv \frac{m\, \re^{B}A}{2\pi}\tilde{\rho},\\
e &=
\frac{m^2 \, \re^{2B} A^2}{2\pi}\left(1+\sum_{m=1}^\infty\frac{1}{B^m}\sum_{s=0}^{m-1}
2^{s+1}Q_{s,m-1-s}\right)\equiv\frac{m^2 \, \re^{2B} A^2}{2\pi}\tilde{e}.
\end{align}
As in the bosonic models, the unknown coefficients can be obtained by matching the bulk and edge answers. One obtains in this way the 
perturbative expansion
\begin{equation}
\label{GN-series}
\ba
4 \frac{\tilde{e}}{\tilde{\rho}^2}&=
1+\alpha  \Delta +\frac{1}{2} \alpha ^2 \Delta  (\Delta +2)
-\frac{1}{2} \alpha ^3 ((\Delta -3) \Delta )\\
&+\frac{1}{8} \alpha ^4 \Delta  \left(\Delta ^3 (1-24 \zeta (3))+42 \Delta ^2 \zeta (3)-\Delta  (21 \zeta (3)+25)+24\right)
\\
&+\frac{1}{12} \alpha ^5 \Delta  \left(120 \Delta ^4 \zeta (3)+\Delta ^3 (7-354 \zeta (3))+\Delta ^2 (357 \zeta (3)+43)-18 \Delta  (7 \zeta (3)+8)+90\right)
\\
& + \CO(\alpha^6). 
\ea
\end{equation}
We have calculated analytically the first $45$ terms of this expansion. It is also possible to analyze the integral equation (\ref{chi-ie}) in a $1/N$ expansion, as already noted in \cite{fnw1,fnw2}. In this way one can obtain all-order results in $\alpha$ for the contribution 
of order $\Delta$ to the above perturbative expansion. This is explained in Appendix \ref{gn-largen}. The result (\ref{largen-e}) provides a further test of (\ref{GN-series}).

\sectiono{Evidence for renormalons}

\label{ren-sec} 
\subsection{Large order behavior from renormalons}

In many asymptotically free theories, renormalons are expected to dominate the large order behavior of conventional perturbation theory. However, as we mentioned in the Introduction, 
diagrammatic arguments are not fully conclusive and it is useful to have another type of reasoning which leads to a precise expectation 
for the large order behavior of perturbation theory. One such argument is provided by the connection between IR renormalons and the OPE, first pointed out by 
Parisi \cite{parisi2}. Let us briefly review this argument (see e.g. \cite{beneke,bly,mmbook}). A generic observable in an asymptotically free QFT can be written as the sum of a perturbative and a non-perturbative contribution:
\be
\label{fg-trans}
\varphi(g)= \varphi_{\rm p}(g) + \varphi_{\rm np}(g),
\ee
where 
\be
\label{pseries}
\varphi_{\rm p}(g)=\sum_{n=0}^{\infty} a_n g^{2n}
\ee
is the perturbative series, and $\varphi_{\rm np}(g)$ is typically exponentially small in the coupling constant $g$. In the terminology of the theory of 
resurgence (see \cite{mmlargen, mmbook, abs} for reviews) we say that $\varphi(g)$ is given by a {\it trans-series} with two different 
small parameters, namely $g^2$ and 
\be
\label{exp-small}
\re^{-A/g^2},
\ee
 where $A$ is an appropriate constant. In some cases, the observable $\varphi(g)$ can be 
studied with an OPE, and this determines the form of $\varphi_{\rm np}(g)$\footnote{In this sense, as pointed out in \cite{shifman}, the OPE gives a physical construction of 
the trans-series (\ref{fg-trans}).}. The OPE will involve contributions of a series of operators $\CO_i$, of dimension $d_i$. Let us focus in the following on the contribution 
of a single operator $\CO$ of dimension $d$. It is of the form, 
\be
\label{np-d}
\varphi_{\rm np}(g) ={1\over Q^d} \langle \CO\rangle_{\mu} C(Q/\mu, g), 
\ee
where $Q$ is an external scale (it could be the external momentum in an Adler function, or the chemical potential $h$ in the situation considered 
in this paper). In (\ref{np-d}) we have indicated explicitly the dependence on the renormalization scale $\mu$, and $C(Q/\mu, g)$ can be computed from 
perturbation theory. Since both $\varphi(g)$ and $\varphi_{\rm p}(g)$ are separately RG-invariant, the same 
must happen to $\varphi_{\rm np}(g)$. Using standard RG arguments (see e.g. \cite{beneke,bly}), and evaluating the non-perturbative correction at $\mu=Q$, we find 
\be
\label{pnprg}
\ba
\varphi_{\rm np}(g) &= C\left(1, g(Q) \right)  \left(\beta_0 g^2 (Q) \right)^{-d \beta_1/(2 \beta_0^2)}
\exp \left(- {d \over 2 \beta_0  g^2 (Q)}\right) \exp \left( -\int^{g(Q)}_{g_0} {\gamma(x) \over \beta(x)} \rd x \right)\\
& \qquad \times  \exp\left( -d \int_{0}^{g(Q)} \left\{  {1\over \beta(x)}+{1\over \beta_0 x^3} -{\beta_1 \over \beta_0^2 x} \right\} \rd x \right),
\ea
\ee
where
\be
\gamma(g)=\gamma^{(1)} g^2+ \cdots
\ee
 is the anomalous dimension of $\CO$, and our convention for the $\beta$ function is as in (\ref{betaf}). In (\ref{pnprg}), $g_0$ is a reference coupling. At leading order in $g$ we find, 
\be
\label{condst-lo}
\varphi_{\rm np}(g) = C\left(1, g(Q) \right) \left( g^2 (Q) \right)^{-\delta}
\exp \left(- {d \over 2 \beta_0  g^2 (Q)}\right) \left( 1 + \CO(g^2)\right), 
\ee
where 
\be
\delta={d \beta_1 \over 2 \beta_0^2} -{\gamma^{(1)} \over 2 \beta_0}. 
\ee
In (\ref{pnprg}) and (\ref{condst-lo}) we have absorbed overall constants in $C\left(1, g(Q) \right)$. As anticipated, $\varphi_{\rm np}(g)$ involves an exponentially small parameter of the form (\ref{exp-small}), where 
\be
\label{Aren}
A={d \over 2 \beta_0}. 
\ee
Let us assume that 
\be
\label{cseries}
C(1, g)= c_{n_0} \left(g^2\right)^{n_0} \left(1+ \CO(g^2)\right), 
\ee
where $n_0$ is a non-negative integer. By a standard argument \cite{mmlargen,mmbook, abs}, the exponentially small 
correction (\ref{condst-lo}) due to the condensate of $\CO$ gives the following contribution to the 
large order behavior of the coefficients $a_n$ appearing in (\ref{pseries}): 
\be
\label{lob}
a_n \sim A ^{-n-b^+} \Gamma\left(n+ b^+ \right), 
\ee
where
\be
\label{bcoeff}
b^+=\delta-n_0={d \beta_1 \over 2 \beta_0^2} -{\gamma^{(1)} \over 2 \beta_0}-n_0. 
\ee
An equivalent statement of this relation can be obtained by considering the Borel transform of $\varphi_{\rm p}(g)$, defined as
\be
\widehat \varphi_{\rm p}(\zeta)= \sum_{n \ge 0} {a_n \over n!} \zeta^n. 
\ee
Then, the large order behavior (\ref{lob}) means that each operator of dimension $d$ appearing in the OPE gives a singularity in the Borel plane located at $\zeta=A$. 
This is usually called an IR renormalon singularity. 

Note that the location of the IR renormalon singularities gives information on the one-loop coefficient of the beta function, and on the operators contributing to the OPE. 
In addition, the coefficient $b^+$ appearing in (\ref{bcoeff}), which determines the next-to-leading correction to the leading asymptotics, has information on the two-loop 
coefficient $\beta_1$ of the beta function and on the anomalous dimension of the corresponding operator.  

There is however another source of large order behavior in QFT: the UV renormalons. In the case of asympotically free theories, 
they lead to terms in the trans-series with {\it positive} exponents. These terms can be related to operators of dimension $d+D$, where $D$ is the dimension 
of spacetime \cite{parisi1,benekeUV,beneke}. Their contribution to the large order behavior is of the form 
\be
(-1)^{n+1} A ^{-n-b^-} \Gamma\left(n+ b^- \right). 
\ee
Here, $A$ is as in (\ref{Aren}), and 
\be
\label{bmcoeff}
b^-=-{d \beta_1 \over 2 \beta_0^2} +{\gamma^{(1)} \over 2 \beta_0}-m_0, 
\ee
where $m_0$ is a non-negative integer. 

In general, in an asymptotically free theory, we will have both IR and UV renormalons. Let us label the operators leading to IR renormalons by the indices $i \in \CI_{\rm IR}$, and 
the operators leading to UV renormalons by $j \in \CJ_{\rm UV}$. Then, the perturbative series has to be extended to a general trans-series with exponential terms of the form 
\be
\label{ts-uvir}
\ba
& \sum_{i \in \CI_{\rm IR}}   C^+_i \left(g^2 \right)^{-b^+_i} \exp \left(- {d_i \over 2 \beta_0  g^2}\right) \left( 1 + \CO(g)\right) \\
&+ \sum_{j \in \CJ_{\rm UV}} C^-_j \left(g^2\right)^{-b^-_j} \exp \left( {d_j \over 2 \beta_0  g^2}\right) \left( 1 + \CO(g)\right).
\ea
\ee
Here, $b_i^+$, $b_j^-$ are given in (\ref{bcoeff}) and (\ref{bmcoeff}), respectively, where we set $d=d_{i,j}$ and $\gamma^{(1)}= \gamma_{i,j}^{(1)}$. 
As a consequence, we find the following large order behavior for the perturbative series:
\be
\ba
a_n &\sim {1\over 2 \pi} \sum_{i \in \CI_{\rm IR}} C^+_i A_i ^{-n-b^+_i} \Gamma\left(n+ b^+_i \right)\\
& + {1\over 2 \pi} \sum_{j \in \CJ_{\rm UV}} 
 C^-_j (-1)^{n+1} A_j^{-n-b^-_j} \Gamma\left(n+ b^-_j \right),  \qquad n \gg 1, 
 \ea
\ee
where
\be
A_{i,j}= {d_{i,j} \over 2 \beta_0}. 
\ee
Here, we have restricted ourselves to the next-to-leading order for the asymptotic expansion. 
Further corrections can be also studied (see e.g. \cite{beneke}), but we will not 
consider them in this paper.

\subsection{Testing renormalon predictions}

The leading large order behavior of the perturbative series will be determined by the IR and UV renormalons with the smallest possible value of $d_i$. In 
two dimensions, there is an IR singularity in the Borel plane at 
\be
\zeta=\beta_0^{-1},
\ee
which corresponds to condensates of dimension $d=2$. There is also an 
UV singularity at $\zeta=-\beta_0^{-1}$, due to operators of dimension $4$ 
(see e.g. the discussion of \cite{david2} on the non-linear sigma model). Let us label the dimension $d=2$ 
 operators contributing to the IR singularity by $i\in \CI_{\rm  IR}^{(2)}$, and the dimension $d=4$ operators contributing to the 
 UV singularity by $j \in \CJ_{\rm UV}^{(2)}$. If we define
\be
\label{norm-version}
c_n = \beta_0^{-n} a_n, 
\ee
we obtain the following renormalon prediction for the large order behavior:
\be
\label{cn-lo}
c_n \sim  \sum_{i \in \CI_{\rm IR}^{(2)}}   C^+_i  \Gamma\left(n+ b^+_i \right) + \sum_{j \in \CJ_{\rm UV}^{(2)}}  C^-_j (-1)^{n+1} \Gamma\left(n+ b^-_j \right), \qquad n  \gg 1. 
\ee
In this formula we have redefined the overall constants to absorb the factor $(2 \pi)^{-1}$. 
Both the UV and the IR renormalon contribute to (\ref{cn-lo}), and in order to test 
this prediction we have to disentangle their contribution. One obvious consequence of the presence of 
both singularities in the Borel plane is that odd and even terms of the $c_n$ series have different large order behavior. We have 
\be
\ba
c_{2k} & \sim \sum_{i \in \CI_{\rm IR }^{(2)}} C^+_i  \Gamma\left(2k+ b^+_i \right) -
 \sum_{j \in \CJ_{\rm UV}^{(2)}} C^-_j \Gamma\left(2k+ b^-_j \right), \\
c_{2k-1} & \sim   \sum_{i \in \CI_{\rm IR }^{(2)}}   C^+_i  \Gamma\left(2k-1+ b^+_i \right) +
\sum_{j \in \CJ_{\rm UV}^{(2)}}  C^-_j \Gamma\left(2k-1+ b^-_j \right), 
\ea
\ee
where $k \gg 1$. Let us now introduce the auxiliary series:
\be
f_k= {c_{2k} \over \Gamma(2k+1)}, \qquad g_k= {c_{2k-1} \over \Gamma(2k)}. 
\ee
Then, we have the large order behavior, 
\be
\ba
f_k &\sim  \sum_{i \in \CI_{\rm IR }^{(2)}}    C^+_i (2k)^{b^+_i -1} \left(1+ \CO(k^{-1})\right)-
 \sum_{j \in \CJ_{\rm UV}^{(2)}} C^-_j (2k)^{b^-_j -1}  \left(1+ \CO(k^{-1})\right),\\
 g_k &\sim  \sum_{i \in \CI_{\rm IR }^{(2)}}   C^+_i (2k)^{b^+_i -1} \left(1+ \CO(k^{-1})\right)+
\sum_{j \in \CJ_{\rm UV}^{(2)}}   C^-_j (2k)^{b^-_j -1}  \left(1+ \CO(k^{-1})\right)
\ea
\ee
for $k \gg 1$. If we define 
\be
\CS_k=  f_k+ g_k , \qquad {\cal D}_k =g_k- f_k, 
\ee
we conclude that 
\be
\label{IRas}
\CS_k \sim  2 \sum_{i \in \CI_{\rm IR }^{(2)}}  C^+_i (2k)^{b^+_i -1} \left(1+ \CO(k^{-1})\right), \qquad k \gg 1, 
\ee
while 
\be
 \label{UVas}
{\cal D}_k\sim  2  \sum_{j \in \CJ_{\rm UV}^{(2)}} C^-_j (2k)^{b^-_j -1} \left(1+ \CO(k^{-1})\right), \qquad k \gg 1. 
\ee
Therefore, the first sequence is sensitive to the first IR renormalon singularity, while the second sequence is sensitive to the first UV renormalon singularity. 

We would like to test the above expectations from the theory of renormalons, against the large order behavior of the perturbative series 
that we have calculated above. Although the arguments we have presented are derived for observables in which the OPE can be used, we will assume that they also control 
the ground state energy in the presence of an external field studied in this paper. We first note that in our perturbative series the coupling 
is defined by (\ref{def_agen}) and (\ref{alpha-GN}) in the bosonic and GN models, respectively. 
In both cases we have 
\be
\alpha \sim 2 \beta_0 \overline g^2.  
\ee
As discussed in \cite{beneke-lo,beneke}, such redefinitions of the coupling constant change the location 
of the singularity in the Borel plane by the overall factor $2 \beta_0$. This means 
the leading IR and UV singularities will be at $\zeta=\pm 2$. However, they do not change the values of 
the coefficients $b_i^+$, $b_j^-$. In particular, the normalized 
version of the series (\ref{norm-version}) involves multiplying by $2^n$ the coefficients of the perturbative series 
obtained in section \ref{pseries-sec}. 

We will mostly focus on the 
IR renormalon singularity. This is because it involves dimension $2$ operators, and there is only a small 
number of these. A detailed study of the UV renormalon singularity requires all the dimension $4$ operators. 
 For example, in the non-linear sigma model there is only one dimension $2$ operator, but five dimension 
 $4$ operators \cite{blgzj,blgzj2}. To extract information about the leading IR singularity, we note that the leading large order behavior of $\CS_k$ is governed by 
the largest values of the $b_i^+$, which we will denote by $b_*^+$. This coefficient can be extracted from the auxiliary sequence
\be
\label{aux-sig}
\sigma_k = k \left( \log\left( \CS_{k+1} \right)-\log\left( \CS_{k} \right) \right), 
\ee
which behaves as 
\be
\label{aux-sig-b}
\sigma_k \sim b_*^+-1 + \CO\left({1\over k} \right), \qquad k \gg 1. 
\ee

Finally, we note that the perturbative series for the bosonic models start with $\alpha$, and not with $\alpha^0$. 
This simply amounts to a redefinition of the coefficients as follows:
for bosonic models we choose our observable for \eqref{fg-trans} to be
\begin{equation}
\varphi_{\text{bosonic}} = \alpha^{-1} \frac{\tilde{e}}{B \tilde{\rho}^2} \,,
\end{equation}
while for the Gross-Neveu model we choose
\begin{equation}
\varphi_{\text{GN}} = 4 \frac{\tilde{e}}{ \tilde{\rho}^2} \,.
\end{equation}
With this choice in both cases we have
\begin{equation}
\varphi_p(\alpha) = \sum_{n=0}^\infty a_n \alpha^n = 1 + \CO(\alpha),
\end{equation}
as can be seen from \eqref{on-ps}, \eqref{susyon-ps}, \eqref{pcf-ps}, \eqref{efkw} and \eqref{GN-series}.
We will now 
present a discussion of the four models we consider in this paper.


\begin{figure}
\center
\includegraphics[height=4cm]{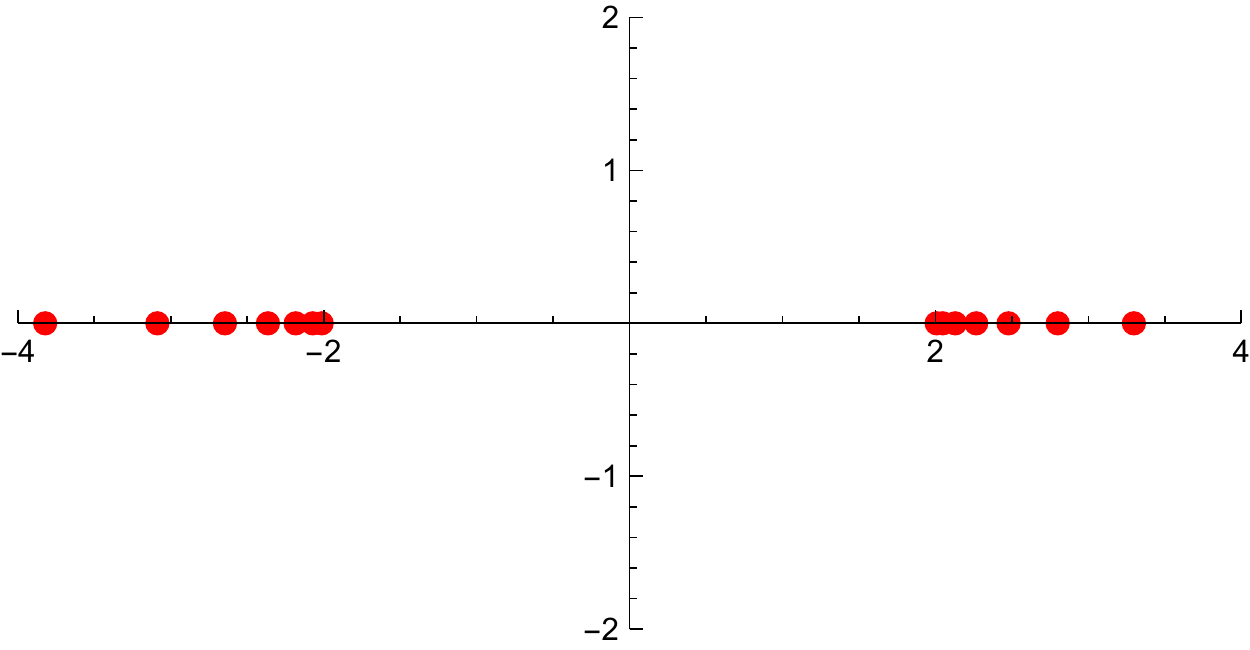}
\caption{In this figure we show the poles of the Borel--Pad\'e transform of the first $44$ terms of the perturbative series (\ref{on-ps}) for the $O(N)$ sigma model with $N=4$. We 
consider the Pad\'e approximant $(21, 21)$. The poles accumulate in two branch cuts starting at the singularities $\zeta=\pm 2$, corresponding 
to the first IR and UV renormalons, respectively.}
\label{poles-on}
\end{figure}

\begin{figure}
\center
\includegraphics[height=4cm]{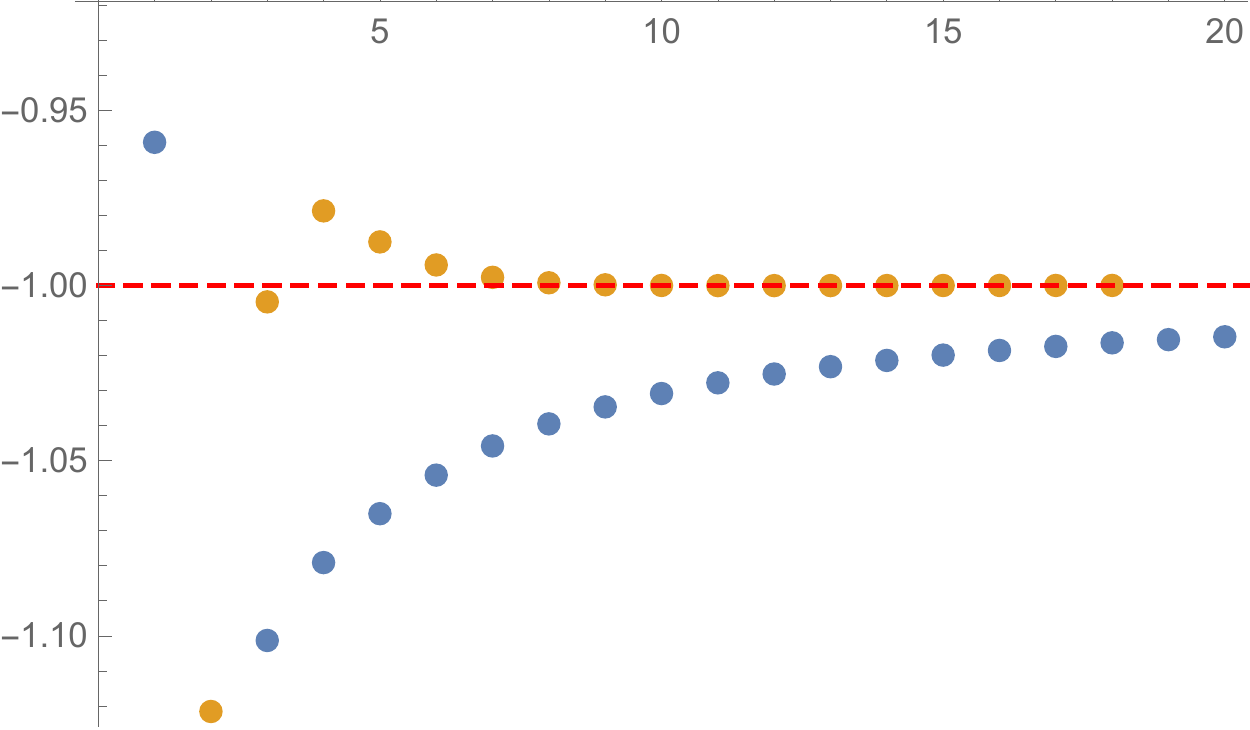} \qquad \qquad \includegraphics[height=4cm]{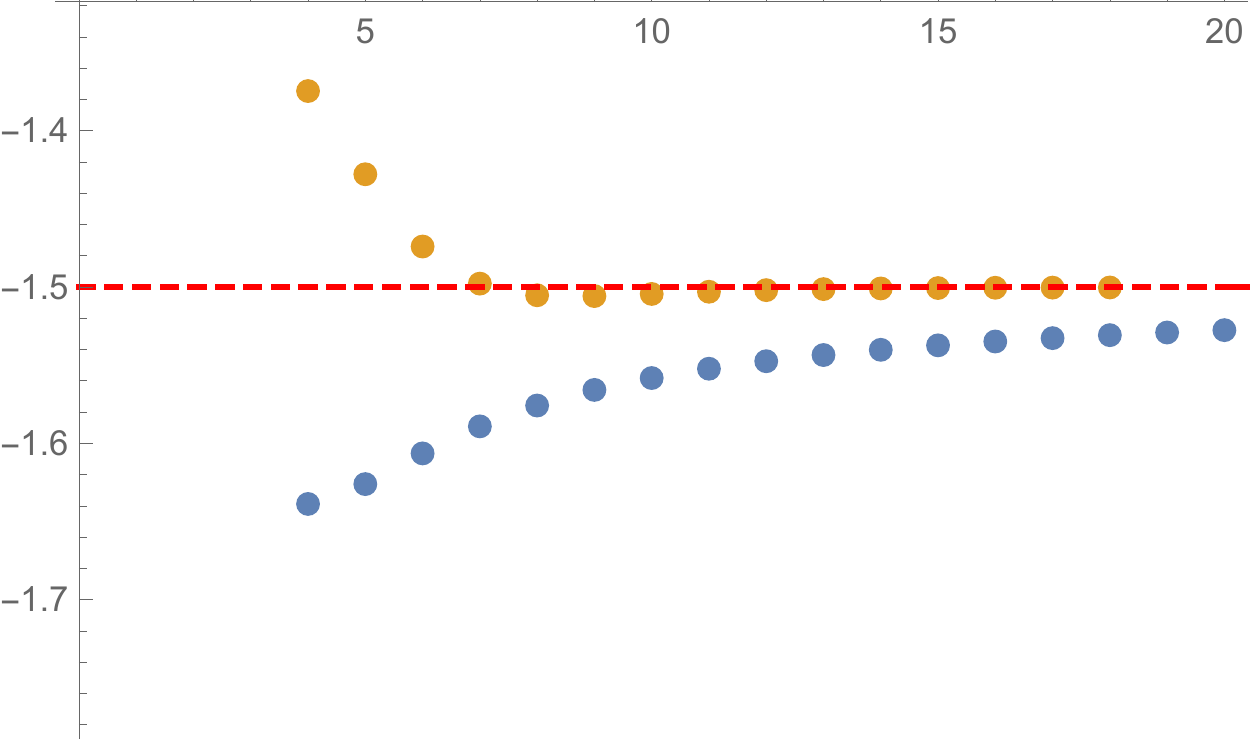}
\caption{The figure in the left (respectively, right) shows the sequence (\ref{aux-sig}) and its second Richardson transform for the $O(N)$ sigma model, where $N=4$ and $N=6$, respectively. The horizontal dashed line is the expected value $2 \Delta-2$.}
\label{on-bs}
\end{figure}

(i) {\it Non-linear $O(N)$ sigma model}. A preliminary analysis of the large order behavior of the perturbative series (\ref{on-ps}) model was performed in \cite{volin}. 
A direct test of the presence of singularities in the Borel plane can be simply done by plotting the 
poles of the Borel--Pad\'e transform of the perturbative series (\ref{on-ps}). As we see in \figref{poles-on} in the case of $N=4$, 
these poles accumulate along branch cuts in the positive and the negative real axes. The location of the branch points is clearly at $\zeta=\pm 2$, signaling the first IR and UV renormalons. The first IR renormalon corresponds to the operator of dimension $2$ given by (see e.g. \cite{david2})
\be
\CO= \partial_\mu {\boldsymbol{S}} \cdot  \partial^\mu {\boldsymbol{S}} . 
\ee
This is the analogue of the gluon operator in Yang--Mills theory. The anomalous dimension of the operator given by the Lagrangian density is
closely related to the beta function of the coupling constant (see e.g. \cite{robertson, gr}). One finds, 
\be
\label{nlsm-g}
\gamma(g)= -{2\over g} \beta(g), 
\ee
so that 
\be
\label{nlsm-g1}
\gamma^{(1)}=2 \beta_0. 
\ee
In particular, in the asymptotics (\ref{IRas}) there is a single term with 
\be
\label{bnlsm}
b^+_*={\beta_1 \over  \beta_0^2}-{\gamma^{(1)} \over 2 \beta_0}-n_0=2 \Delta-1-n_0. 
\ee
In order to test this prediction, we use our data for the perturbative series to construct the sequence (\ref{aux-sig}), and we study its asymptotic behavior for 
different values of $N$. To remove tails and to accelerate the convergence to $b^+_*$, we can use Richardson transforms (see e.g. \cite{mmbook}). Our results vindicate 
the result (\ref{aux-sig-b}) with $b_*^+$ given in (\ref{bnlsm}), and $n_0=0$. Note that, by consistency, this provides a test not only of the value of $b_*^+$, but also of 
the assumptions leading to the prediction (\ref{aux-sig}), namely the existence of an IR renormalon singularity 
at $\zeta=2$, and the factorial growth of the sequence. As an example, in \figref{on-bs} we show the sequence (\ref{aux-sig}) and its second Richardson transform for $N=4$ (left) and $N=6$ (right). After two Richardson transforms, the sequences approach the value $b_*^+-1=2 \Delta-2$ with an error of $10^{-5}$ and $10^{-4}$, respectively. We have 
tested this for many other values of $N$. The behavior persists even for rational values of $N$, although the tests become less precise as $N$ becomes large. 

\begin{figure}
\center
\includegraphics[height=4cm]{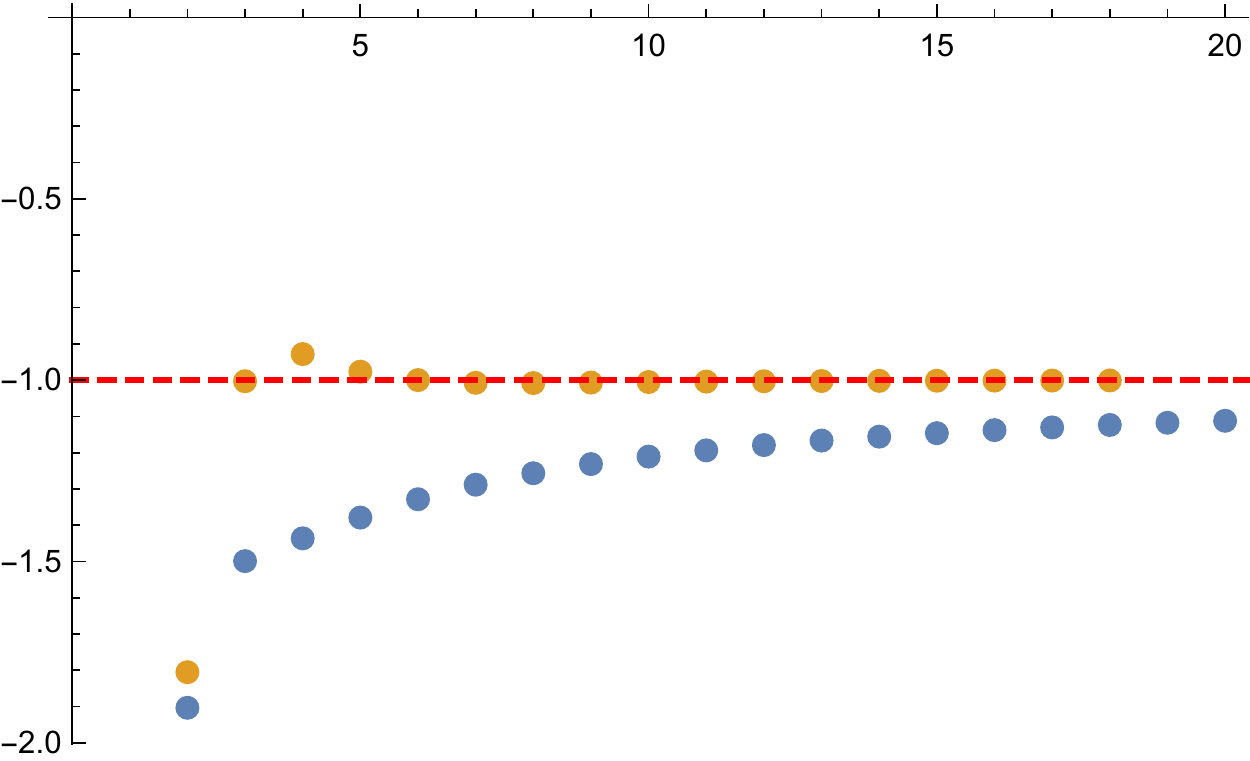} \qquad \qquad \includegraphics[height=4cm]{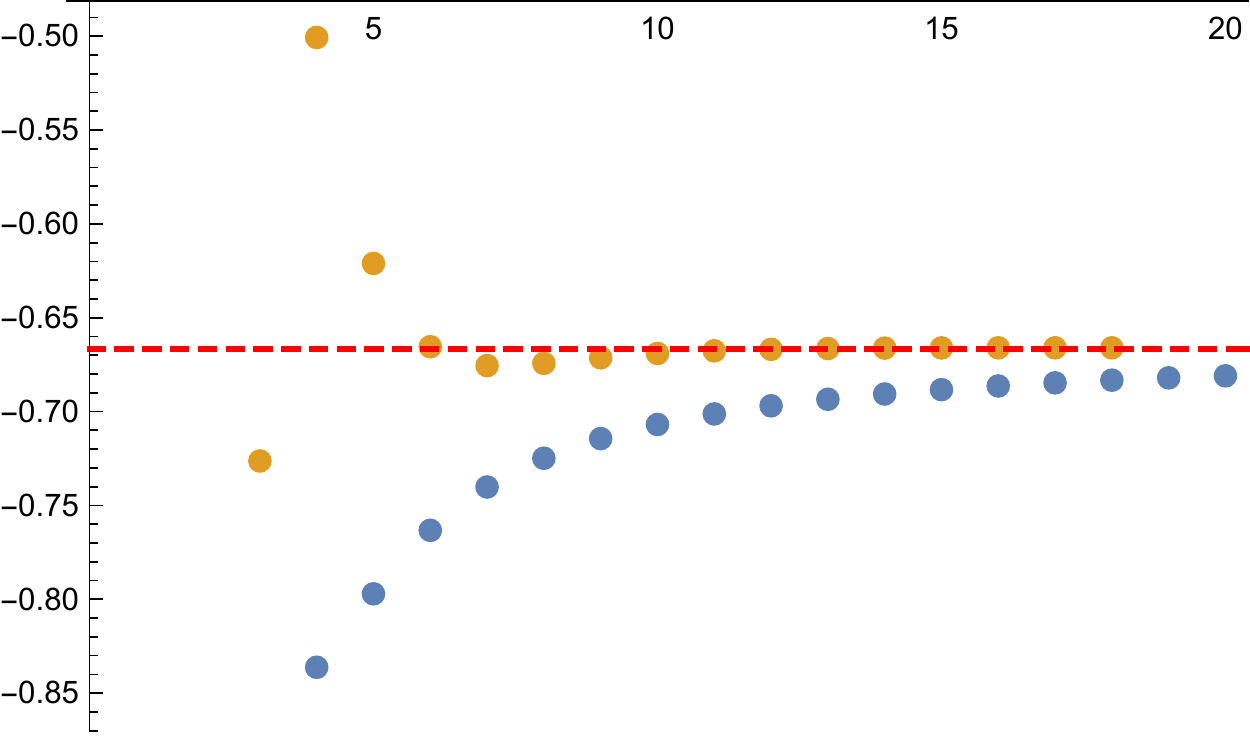}
\caption{The figure in the left (respectively, right) shows the analogue of the sequence (\ref{aux-sig}) with ${\cal D}_k$ instead of ${\cal S}_k$, as well as its second Richardson transform, for the $O(N)$ sigma model with $N=4$ and $N=5$, respectively. The horizontal dashed line is the expected value $-2 \Delta$.}
\label{bm-fig}
\end{figure}

It is also possible in this case to analyze the UV renormalon singularity at $\zeta=-2$ in some detail, by looking at an auxiliary sequence similar to (\ref{aux-sig-b}), but where we replace $\CS_k$ by ${\cal D}_k$. Our results indicate that the asymptotic behavior is controlled by an exponent
\be
b_*^-=-2 \Delta +1, 
\ee
as we show in two examples in \figref{bm-fig}. On the other hand, the UV renormalon should be associated to operators of dimension $4$. A complete basis for these, involving five different operators, 
was obtained in \cite{blgzj,blgzj2}, and their anomalous dimensions computed. One finds the 
possible values
\be
{\gamma^{(1)} \over 2 \beta_0}= 1, \, 2, \, 1-\Delta, \, 2+\Delta, \, -2 \Delta. 
\ee
The first two values are compatible with our empirical finding (and appropriate values of $m_0$, namely $m_0=0$ and $m_0=1$, respectively). If this picture is correct, our numerical 
result verifies the fact that, in the UV renormalon, the next-to-leading asymptotics (\ref{bmcoeff}) involves the quotient $-\beta_1/(2 \beta_0^2)$ (i.e. with the opposite sign than the IR renormalon). 

\vskip .2cm

(ii) {\it $\CN=1$ non-linear $O(N)$ sigma model}. The behavior of the supersymmetric non-linear sigma model is more complicated. 
The large order asymptotics is dominated by an UV renormalon at $\zeta=-2$ with $b_-^*=1$, and independent of $N$. 
There is however an IR renormalon at $\zeta=2$. 
The easier way to see this is by considering the large $N$ limit. The coefficients in the perturbative series (\ref{susyon-ps}) have the following 
$1/N$ expansion: 
\be
c_n = \Delta c_n ^{(0)} + \CO(\Delta^2), \qquad n \ge 3. 
\ee
It is easy to verify numerically that the coefficients $c_n^{(0)}$ lead to an IR renormalon singularity at $\zeta=2$ with $b_+^*= -1$. 
The behavior of the IR renormalon at finite $N$ is more complicated, since there seems to be an additional contribution to the large order 
asymptotics which has $b_+^*=-2$. To determine in detail these subleading contributions we probably need more terms in the perturbative series. 

In \cite{dsu} it was argued that the leading IR renormalon singularity is absent in a class of correlation 
functions of some supersymmetric theories. In view of the IR renormalon at $\zeta=2$ in the supersymmetric non-linear sigma model, we conclude that the mechanism of \cite{dsu} is not generic and does not apply here.

\vskip .2cm

\begin{figure}
\center
\includegraphics[height=4cm]{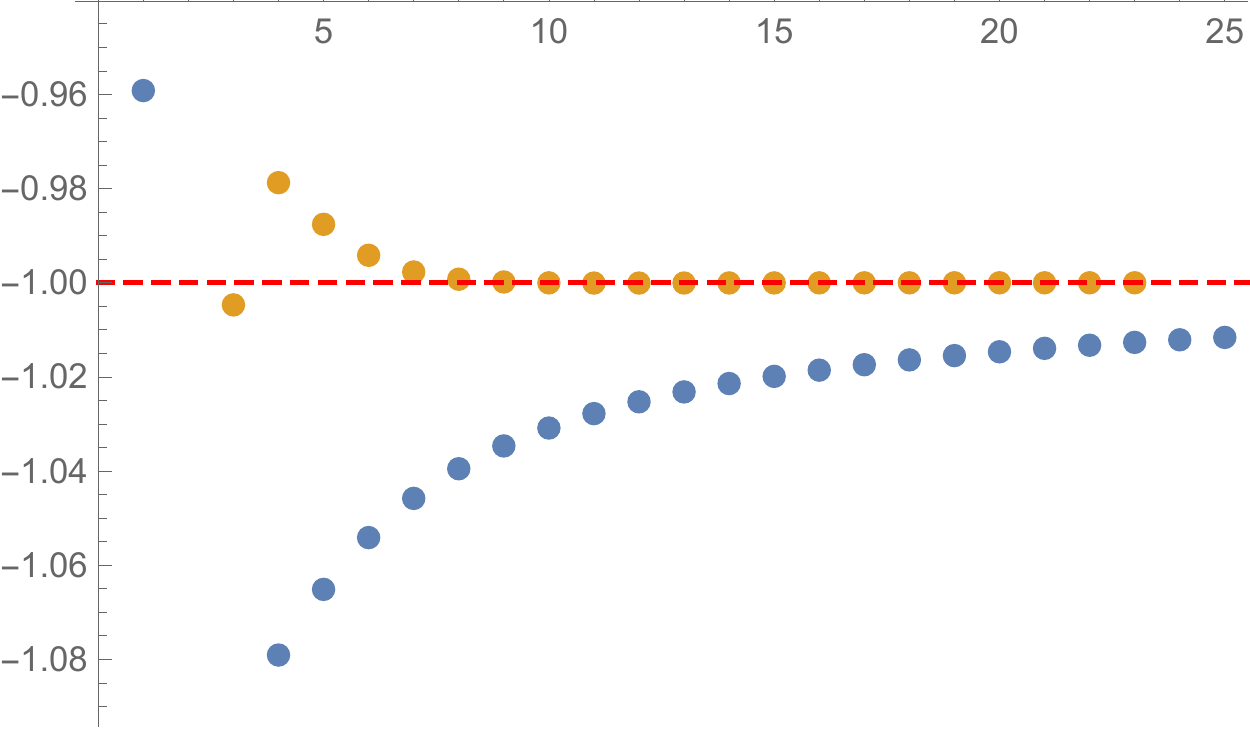} \qquad \qquad \includegraphics[height=4cm]{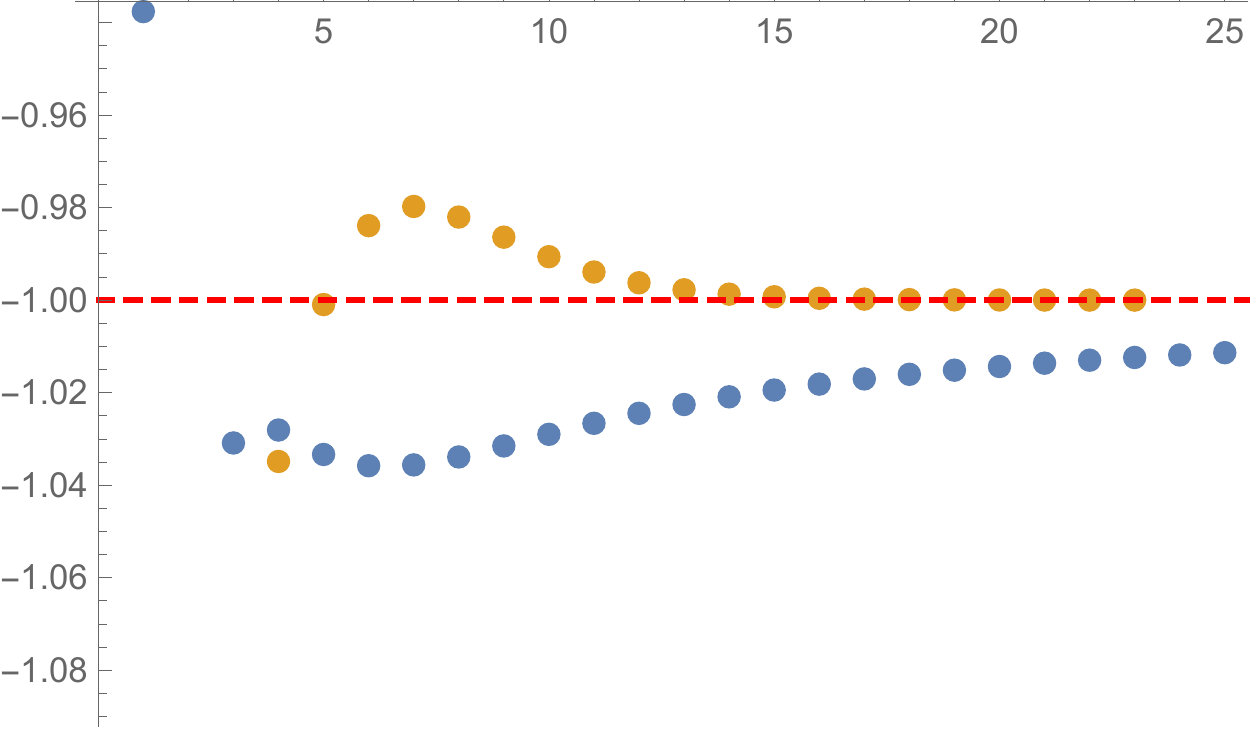}
\caption{The figure in the left (respectively, right) shows the sequence (\ref{aux-sig}) and its second Richardson transform for the $SU(N)$ PCF, where $N=2$ and $N=3$, respectively. The horizontal dashed line is the expected value $-1$.}
\label{pcf-bs}
\end{figure}

(iii) {\it $SU(N)$ principal chiral field}. The PCF is very similar to the non-linear sigma model. The first IR renormalon corresponds to 
the operator of dimension $2$ appearing in the Lagrangian density:
\be
\CO=\tr \left(\partial_\mu \Sigma \, \partial^\mu \Sigma^\dagger \right).
\ee
Its anomalous dimension is given again by the formulae (\ref{nlsm-g}), (\ref{nlsm-g1}). There is a single term in (\ref{IRas}) with 
\be
\label{bppcf}
b^+={\beta_1 \over  \beta_0^2}-{\gamma^{(1)} \over 2 \beta_0}-n_0=-n_0. 
\ee
Note in particular that this is independent of $N$. Our numerical calculations, using for example the poles of the Borel--Pad\'e transform, 
indicate clearly the presence of IR and UV renormalons at $\zeta=\pm 2$. Moreover, they vindicate the value (\ref{bppcf}) with $n_0=0$. In \figref{pcf-bs} we show the sequence (\ref{aux-sig}) and its second Richardson transform for $N=2$ (left) and $N=3$ (right). After two Richardson transforms, the sequences approach the value $b_*^+-1=-1$ with an error of $10^{-5}$ and $10^{-6}$, respectively. As in the $O(N)$ sigma model, this behavior persists for larger values of $N$, although precision decreases. 

\vskip .2cm

\begin{figure}
\center
\includegraphics[height=4cm]{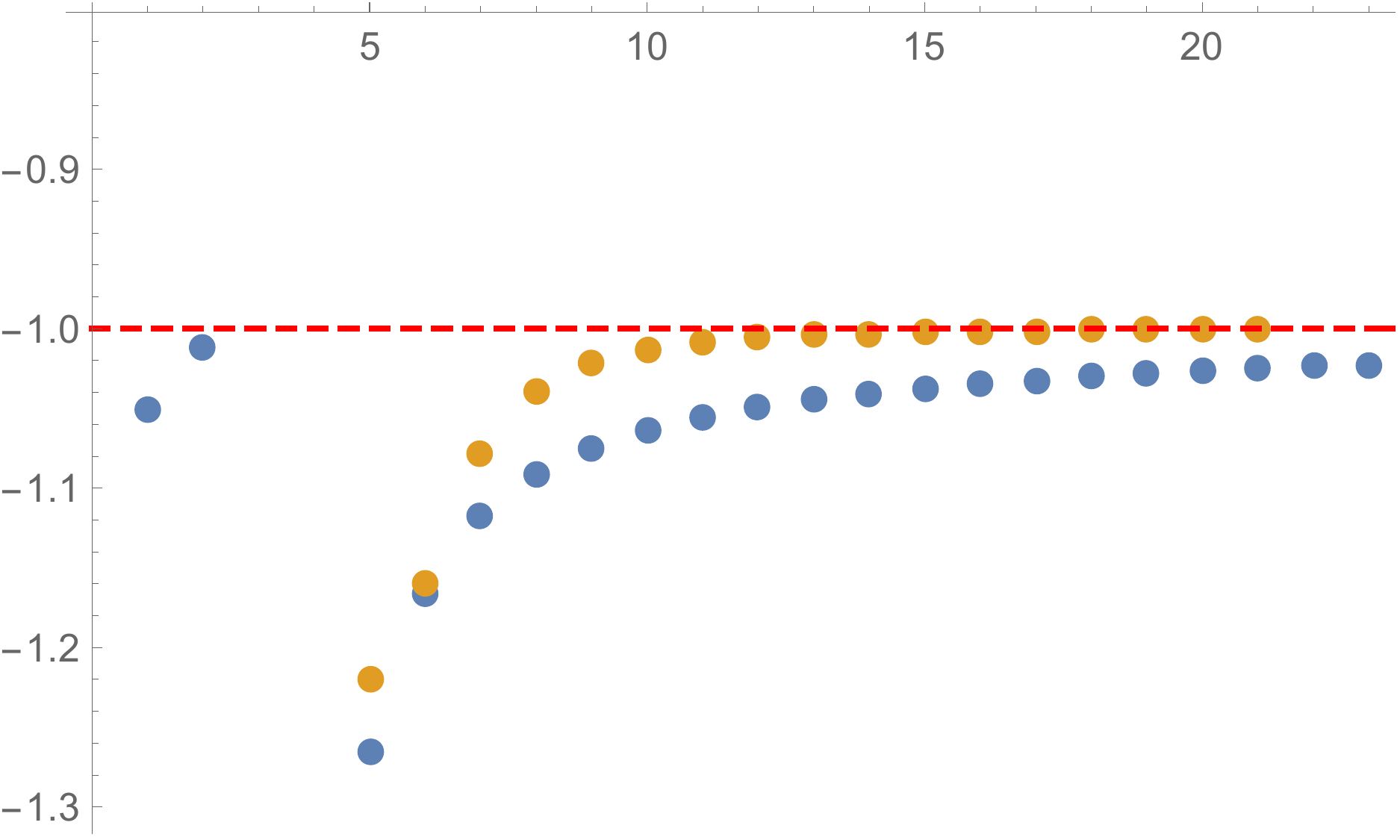} \qquad \qquad \includegraphics[height=4cm]{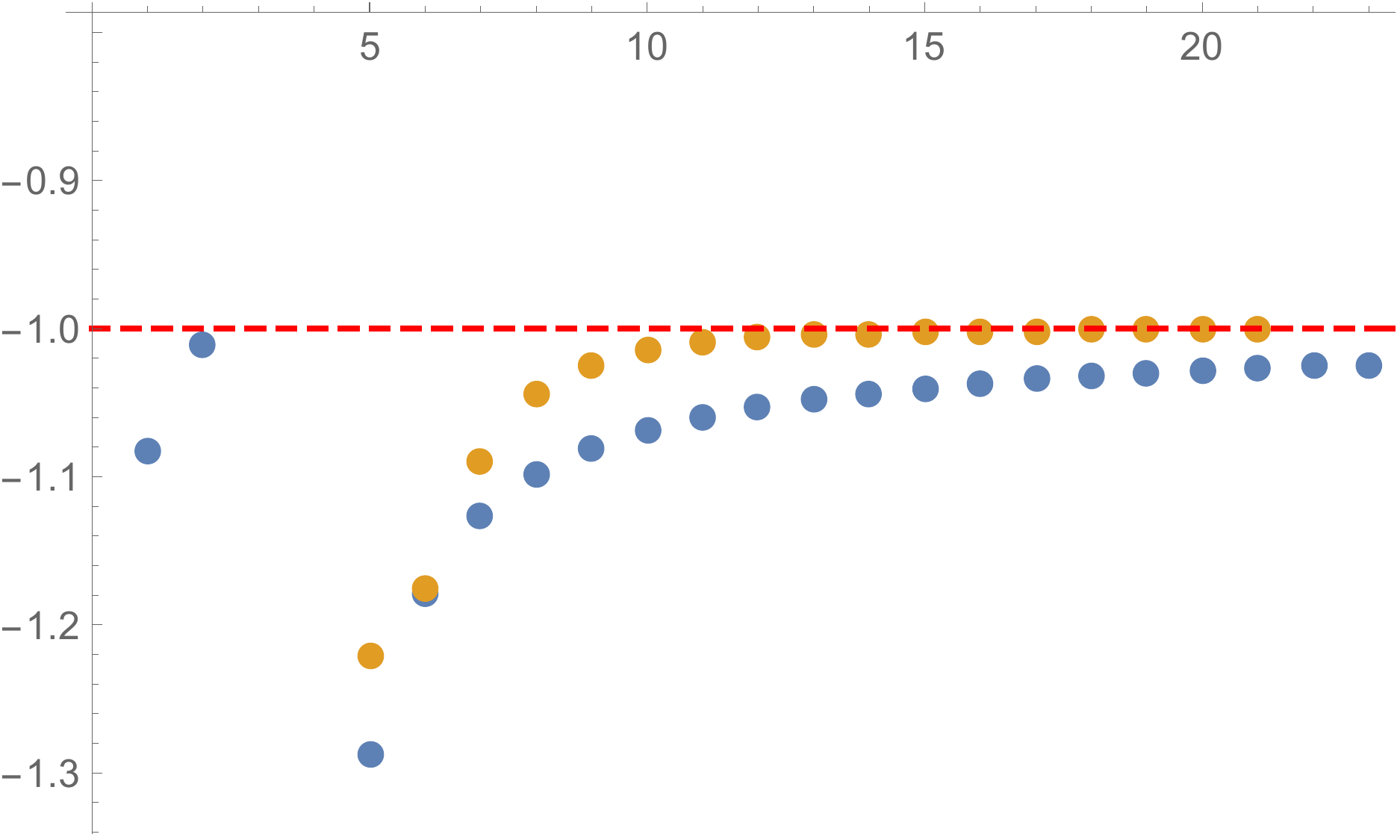}
\caption{The figure in the left (respectively, right) shows the sequence (\ref{aux-sig}) and its second Richardson transform for the $SU(N)$ PCF with FKW conserved charges, where $N=2$ and $N=7$, respectively. The horizontal dashed line is the expected value $-1$.}
\label{fkw-bs}
\end{figure}

(iv) {\it $SU(N)$ principal chiral field with FKW charges}. As one would expect, despite the different perturbative series, the asymptotics of the coefficients remains the same since the renormalon physics should be unaffected by the charge choice. Indeed, as shown in \figref{fkw-bs}, an identical analysis finds $b_*^+-1=-1$ with an error of order $10^{-4}$ for $N=2$ and $N=7$, for example.

\vskip .2cm

\begin{figure}
\center
\includegraphics[height=4cm]{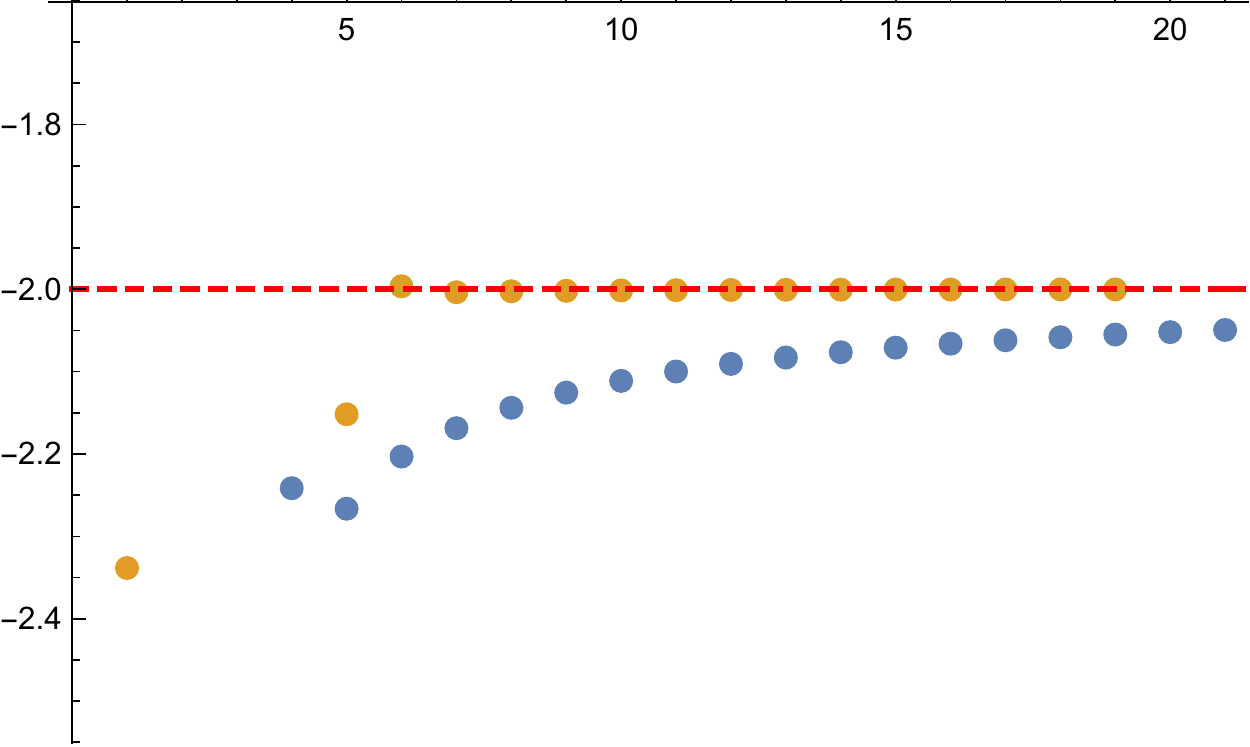} \qquad \qquad \includegraphics[height=4cm]{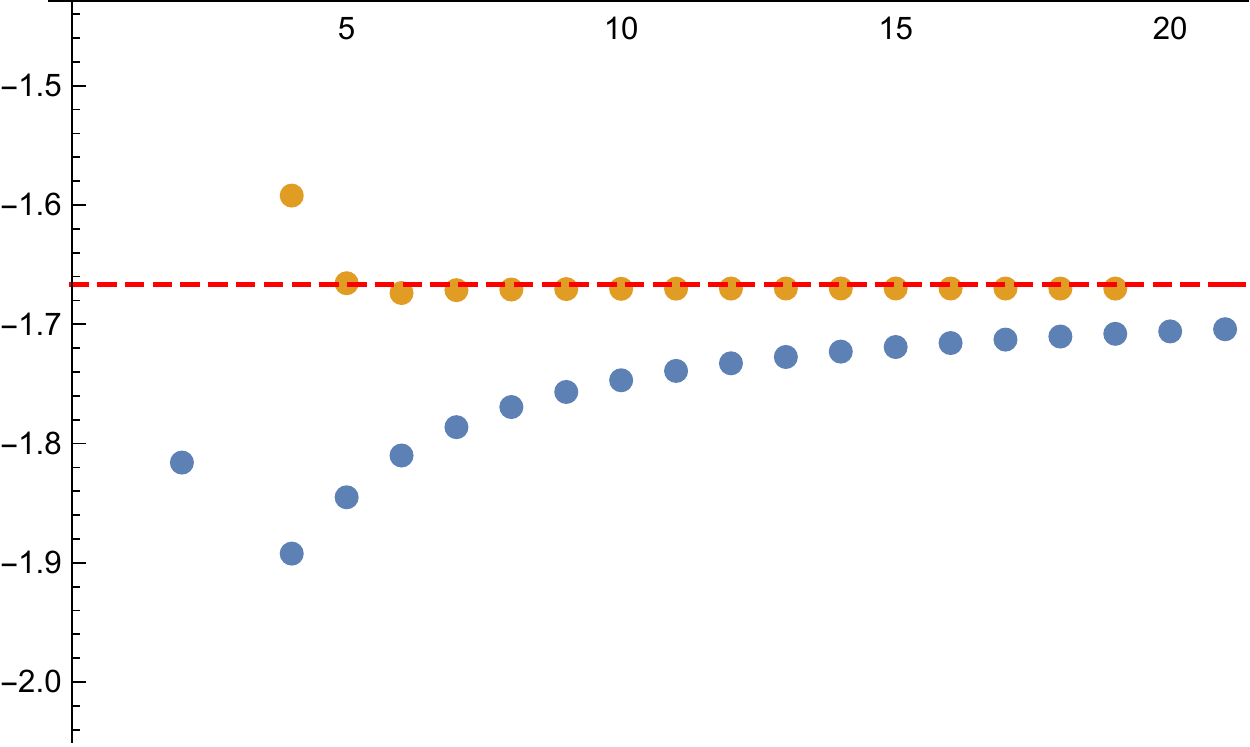}
\caption{The figure in the left (respectively, right) shows the sequence (\ref{aux-sig}) and its second Richardson transform for the $O(N)$ GN model, where $N=4$ and $N=5$, respectively. The horizontal dashed line is the expected value $-1-2\Delta$.}
\label{gn-bs}
\end{figure}

(v) {\it $O(N)$ Gross--Neveu model}. There are now two operators of dimension $2$, corresponding to the two operators in the Lagrangian:
\be
 \CO_1= \left( \overline{\boldsymbol{\chi}} \cdot  \boldsymbol{\chi} \right)^2, \qquad \CO_2= \overline{\boldsymbol{\chi}} \cdot  \gamma^\mu \partial_\mu  \boldsymbol{\chi}. 
\ee
The calculation of their anomalous dimensions can also be easily done, by using e.g. the results of \cite{gracey} on the 
renormalization properties of this model. One finds, at one-loop, 
\be
\gamma^{(1)}_1=-2 \beta_0, \qquad \gamma^{(1)}_2=0. 
\ee
This result agrees with the calculation in \cite{gn-anomalous}. Therefore, in the asymptotics (\ref{IRas}) there will be two terms. In the first one, corresponding to 
the quartic fermion term, one has
\be
\label{b1gn}
b_1^+={\beta_1 \over  \beta_0^2}-{\gamma^{(1)}_1\over 2 \beta_0}-n_{0,1}=-2 \Delta+1-n_{0,1}, 
\ee
while in the second one, corresponding to the kinetic term, 
\be
\label{b2gn}
b_2^+={\beta_1 \over  \beta_0^2}-{\gamma^{(1)}_2 \over 2 \beta_0}-n_{0,2}=-2 \Delta-n_{0,2}.  
\ee
According to our numerical results, the leading term in the asymptotics has $b_*^+=-2 \Delta$. This corresponds to (\ref{b1gn}) with $n_{0,1}=1$ or to (\ref{b2gn}) with $n_{0,2}=0$. 
In \figref{gn-bs} we show the sequence (\ref{aux-sig}), as well as its second Richardson transform, for $N=4$ (left) and $N=5$ (right). The numerical result agrees the expected result 
$b_*^+-1=-1 -2 \Delta$ with an error of $10^{-4}$ and $10^{-3}$, respectively. 

\vskip .2cm

We conclude that the large order behavior of the perturbative series obtained in section \ref{pseries-sec} are indeed 
controlled by renormalons (provided some reasonable assumptions are made on the values of the exponents $n_0$ or $m_0$). 
In particular, we can isolate the contribution from 
IR and UV renormalons separately. This makes it possible to test the next-to-leading contribution of the IR renormalon 
to the asymptotics in the four models considered in this paper, which is given by (\ref{bcoeff}). This value is sensitive to the 
first two coefficients of the beta function. In practice, this is clearly seen in our numerical analysis when we change $N$: in the $O(N)$ sigma model and in the GN model, this coefficient involves $\pm 2/(N-2)$, respectively, while in the PCF it is independent of $N$.

\sectiono{Conclusions}

Since the pioneering work in \cite{pw}, it is well known that in integrable field theories in two dimensions the ground state energy 
can be calculated exactly once a conserved charge is coupled to an external field. Extracting the perturbative series from the 
resulting Bethe ansatz equation turns out to be challenging. In this paper we have built upon \cite{volin, volin-thesis} to obtain the perturbative 
series expansion in a number of integrable field theories. In this way we have been able to provide a direct test of renormalon predictions for the 
large order behavior of these series. 
In particular, we have tested the next-to-leading correction to its asymptotics, involving the first two coefficients of the beta function, as well as the anomalous dimensions of the operators appearing in the OPE. 

Although we believe that our tests of renormalon predictions are convincing, there are various issues that should be clarified. For example, 
we used predictions about the large order behavior based on OPE considerations. Our observable involves perturbing the 
Lagrangian with an integrated conserved current, so it seems reasonable that the OPE can be used 
to establish our working hypothesis, but one should address this point more carefully. Also, in our tests we made some additional assumptions on the 
exponents $n_0$, $m_0$ appearing in (\ref{cseries}) and (\ref{bmcoeff}). It would be interesting to derive these assumptions from first principles. 

Some technical aspects of our analysis could also be improved. For example, although we have produced $40$-$50$ coefficients in the 
different perturbative series, with the current implementation 
of the method there is a computational bottleneck beyond $50$ terms. If we were able to generate many more terms, it is very likely that we could improve 
our tests of IR renormalon behavior, which become less precise as $N$ becomes large. Clearly, our methods could be straightforwardly 
applied to other integrable field theories.

There are more ambitious research directions open by our results. In this paper we have extracted the 
perturbative series from the Bethe ansatz equation, but we have neglected exponentially 
small terms at large $B$. In principle one could incorporate these terms in a systematic way to obtain 
the {\it full} trans-series (\ref{ts-uvir}) associated to renormalons. In other words, one 
should be able to express the exact $e$, $\rho$ as Borel--\'Ecalle resummations of suitable trans-series, extracted 
from the integral equation (\ref{chi-ie}). Such a trans-series solution 
would give an enormous wealth of information on the renormalon physics of these field theories, including 
exact values for the condensates (modulo Stokes jumps). 

Another interesting direction is to understand the fate of renormalons after a (twisted) compactification on a circle. 
It has been suggested that, after such a compactification, renormalons disappear as such \cite{tin} and the corresponding singularities can be realized 
semiclassically \cite{argyres-unsal, a-unsal-long, dunne-unsal-cpn, cherman-dorigoni-dunne-unsal, cdu, misumi-1,du-on, misumi-2}  (see \cite{circle} for related work). 
In particular, \cite{cherman-dorigoni-dunne-unsal, cdu, du-on} provide a concrete semiclassical picture in two of the models considered in this paper, 
namely the principal chiral field and the non-linear sigma model. 
It would be interesting to use integrability techniques to compute observables in the twisted compactification of these theories, 
as a function of the compactification radius. In this way one could study in detail the behavior of renormalon singularities and their 
eventual transmutation in semiclassical instanton singularities.

\section*{Acknowledgements}
We would like to thank Matthias Jamin, Ram\'on Miravitllas and Matthias Puhr for useful 
discussions and correspondence. We are specially indebted to Gerald Dunne, Santi Peris and Mithat Unsal for their comments after a detailed 
reading of the manuscript. This work has been supported in part by the Fonds National Suisse, 
subsidies 200021-156995 and 200020-141329, by the NCCR 51NF40-141869 ``The Mathematics of Physics'' (SwissMAP), and by the ERC Synergy Grant ``ReNewQuantum". 

\appendix

\sectiono{Large $N$ expansion in the Gross--Neveu model}
\label{gn-largen}
All the models we have considered in this paper can be studied in the large $N$ expansion. In principle this can be done directly in the integral equation (\ref{chi-ie}). However, 
in the bosonic models, the $1/N$ expansion of the kernel leads to a singular function 
at subleading order. In the case of the GN model, as already noted in \cite{fnw1}, the kernel does admit
 a regular $1/N$ expansion of the form
\be
\label{k-expansion}
K(\theta)= \sum_{k \ge 0} \Delta^k K^{(k)}(\theta), 
\ee
where
\be
K^{(1)}(\theta)= {1\over \theta^2} -{\cosh(\theta) \over \sinh^2(\theta)}. 
\ee
This makes it possible to solve the integral equation (\ref{chi-ie}) in a $1/N$ expansion. The expansion (\ref{k-expansion}) leads 
to 
\be
\chi(\theta)= \sum_{k \ge0} \Delta^k \chi^{(k)}(\theta), \qquad \rho= \sum_{k \ge 0} \Delta^k \rho^{(k)}, \qquad e=  \sum_{k \ge 0} \Delta^k e^{(k)}.
\ee
We can compute explicitly 
\be
\ba
\rho^{(0)}&= {m \over \pi} \sinh(B), \\
{2 \pi \over m} \rho^{(1)}&=\int_{-B}^B \int_{-B}^B K^{(1)}(\theta-\theta') \cosh(\theta') \rd \theta \rd \theta'\\
&= 2\cosh(B) \left( \gamma-{\rm Chi}(2B)+ \log(\sinh(B))\right)+ 2\sinh(B) \left(-2B + {\rm Shi}(2B) \right), 
\ea
\ee
where 
\be
{\rm Shi}(z)=\int_0^z {\sinh(t)\over t} \rd t, \qquad {\rm Chi}(z)=\gamma_E + \log(z) + \int_0^z{\cosh(t)-1 \over t} \rd t
\ee
are the sinh and cosh integral functions, respectively (see \cite{chodos} for a similar calculation). We also find, 
\be
\ba
{2 \pi \over m^2}e^{(0)} &=B + \cosh(B) \sinh(B),\\
{2 \pi  \over m^2} e^{(1)}& =\left(2B + \sinh(2B)\right) {\rm Shi}(2B)-2 \cosh^2(B) {\rm Chi}(2B)\\
&+ 2\cosh^2(B) \log\left( \sinh(2B) \right) - 2B \sinh(2B) \\
& +(\gamma_E-1) \cosh(2B) -2B^2+ 1+ \gamma_E. 
\ea
\ee

In order to make contact with the perturbative results, we first consider the asymptotic expansions of the hyperbolic integral functions at large $B$. We have, for $|z| \gg 1$, 
\be
\ba
{\rm Shi}(z)&\sim \cosh(z)\Sigma_1(z) + \sinh(z) \Sigma_2(z), \\
{\rm Chi}(z)&\sim \sinh(z)\Sigma_1(z) + \cosh(z) \Sigma_2(z), 
\ea
\ee
where
\be
\Sigma_1(z) ={1\over z} \sum_{k \ge 0} {(2k)! \over z^{2k}}, \qquad \Sigma_2 (z) ={1\over z} \sum_{k \ge 0} {(2k+1)! \over z^{2k+1}}. 
\ee
If we neglect exponentially small terms in $\re^{-B}$, we find 
\be
\label{largeb-rho}
\ba
{2 \pi  \re^{-B} \over m} \rho^{(0)}&\sim 1, \\
{2 \pi  \re^{-B} \over m} \rho^{(1)}&\sim  \gamma_E -\log(2) - \sum_{n \ge 0} {n! \over (2B)^{n+1}}, 
\ea
\ee
as well as 
\be
\label{largeb-eps}
\ba
{2 \pi \over m^2}e^{(0)} &\approx {\re^{2B} \over 4},\\
 {2 \pi \over m^2}e^{(1)} &\approx {\re^{2B} \over 2 }\left( \gamma_E -\log(2) + \sum_{n\ge 1} {n \, n!  \over (2B)^{n+1}} \right). 
 \ea
 \ee
From (\ref{largeb-eps}) and (\ref{largeb-rho}) we obtain 
\be
\label{erho-int}
{2 e(\rho) \over \pi \rho^2} = 1+ \Delta \sum_{n \ge 0} {(n+1)! \over 2^n} B^{-n-1}. 
\ee
Now we use (\ref{alpha-GN}) to find the expression of $B$ as a function of $\alpha$. Since the terms proportional to $B^{-n}$ in (\ref{erho-int}) are already proportional to $\Delta$, it is enough 
to calculate $B$ as a function of $\alpha$ at order $\CO(1)$ in the $1/N$ expansion. One finds $B\approx \alpha^{-1}$ at this order, and we conclude that 
\be
\label{largen-e}
\ba
4 { \tilde e(\rho) \over \tilde \rho^2} &= 1+ 2\Delta \sum_{n \ge 1} n!  \left({\alpha \over 2}\right)^{n}+ \CO(\Delta^2)\\
&=1+ \Delta\left( \alpha + \alpha^2 + {3 \over2 } \alpha^3+ 3 \alpha^4  + {15 \over 2} \alpha^5+ \cdots \right)+ \CO(\Delta^2),
\ea
\ee
which is in precise agreement with the results in (\ref{GN-series}). Note that the series diverges factorially, and 
leads to a Borel singularity at $\zeta=2$. Therefore, at large $N$, only the leading IR renormalon singularity survives. Similar all-order results in the coupling constant at leading order in the $1/N$ expansion were obtained in the PCF with an appropriate choice of 
charge, in \cite{fkw1,fkw2}.

\bibliographystyle{JHEP}

\linespread{0.6}
\bibliography{biblio-renormalons}

\end{document}